\documentclass[%
 reprint,
 superscriptaddress,
 amsmath,amssymb,
 aps,
floatfix,
]{revtex4-2}

\usepackage{graphicx}
\usepackage{dcolumn}
\usepackage{bm}
\usepackage[colorlinks=true, allcolors=blue]{hyperref}

\usepackage{hypbmsec} 
\usepackage{subfigure}
\usepackage{braket} 
\usepackage{blkarray} 
\usepackage{float} 
\usepackage{fp} 

\newcommand{\THztoeV}[2][2]{
    \FPset\x{0.00413566553853599}
    \FPset\y{#2}
    \FPmul\z\x\y
    \FPeval\z{round(\z:#1)}
    \z\;\text{eV}
}

\begin{document}

\preprint{APS/123-QED}

\title{\textit{Ab initio} and group theoretical study of properties of the \texorpdfstring{C\textsubscript{2}C\textsubscript{N}}{} carbon trimer defect in \mbox{h-BN}}

\author{Omid Golami}
\affiliation{%
 Institute for Quantum Science and Technology, University of Calgary, Calgary, Alberta, Canada T2N 1N4
}%
\affiliation{%
 Department of Physics and Astronomy, University of Calgary, Calgary, Alberta, Canada T2N 1N4
}%
\author{Kenneth Sharman}
\affiliation{%
 Institute for Quantum Science and Technology, University of Calgary, Calgary, Alberta, Canada T2N 1N4
}%
\affiliation{%
 Department of Physics and Astronomy, University of Calgary, Calgary, Alberta, Canada T2N 1N4
}%
\author{Roohollah Ghobadi}
\affiliation{%
 Institute for Quantum Science and Technology, University of Calgary, Calgary, Alberta, Canada T2N 1N4
}%
\affiliation{%
 Department of Physics and Astronomy, University of Calgary, Calgary, Alberta, Canada T2N 1N4
}%
\author{Stephen C. Wein}
\affiliation{%
 Institute for Quantum Science and Technology, University of Calgary, Calgary, Alberta, Canada T2N 1N4
}%
\affiliation{%
 Department of Physics and Astronomy, University of Calgary, Calgary, Alberta, Canada T2N 1N4
}%
\author{Hadi Zadeh-Haghighi}
\affiliation{%
 Institute for Quantum Science and Technology, University of Calgary, Calgary, Alberta, Canada T2N 1N4
}%
\affiliation{%
 Department of Physics and Astronomy, University of Calgary, Calgary, Alberta, Canada T2N 1N4
}%
\author{Claudia Gomes da Rocha}
\affiliation{%
 Department of Physics and Astronomy, University of Calgary, Calgary, Alberta, Canada T2N 1N4
}%
\author{Dennis R. Salahub}
\affiliation{%
 Institute for Quantum Science and Technology, University of Calgary, Calgary, Alberta, Canada T2N 1N4
}%
\affiliation{%
 Department of Physics and Astronomy, University of Calgary, Calgary, Alberta, Canada T2N 1N4
}%
\affiliation{
 Department of Chemistry, CMS Centre for Molecular Simulation, and Quantum Alberta, University of Calgary, Calgary, Alberta, Canada T2N 1N4
}%
\author{Christoph Simon}%
\affiliation{%
 Institute for Quantum Science and Technology, University of Calgary, Calgary, Alberta, Canada T2N 1N4
}%
\affiliation{%
 Department of Physics and Astronomy, University of Calgary, Calgary, Alberta, Canada T2N 1N4
}%




\date{\today}

\begin{abstract}
    Hexagonal boron nitride (\mbox{h-BN}) is a promising platform for quantum information processing due to its potential to host optically active defects with attractive optical and spin properties. Recent studies suggest that carbon trimers might be the defect responsible for single-photon emission in the visible spectral range in \mbox{h-BN}. In this theoretical study, we combine group theory together with density functional theory (DFT) calculations to predict the properties of the neutral C\textsubscript{2}C\textsubscript{N} carbon trimer defect. We find the multi-electron states of this defect along with possible radiative and non-radiative transitions assisted by the spin-orbit and the spin-spin interactions. We also investigate the Hamiltonian for external magnetic field and ground-state hyperfine interactions. Lastly, we use the results of our investigation in a Lindblad master equation model to predict an optically detected magnetic resonance (ODMR) signal and the $g^2(\tau)$ correlation function. Our findings can have important outcomes in quantum information applications such as quantum repeaters used in quantum networks and quantum sensing.
\end{abstract}

\maketitle


\section{\label{sec:intro} Introduction}

Color centers as solid-state artificial atoms in systems such as diamond, silicon carbide, and Van der Waals materials, have potential applications in quantum technology \cite{awschalom2013quantum, weber2010quantum}. Many of these color centers are single-photon sources and have good spin properties \cite{aharonovich2016solid}. Single-photon emitters (SPE) are a vital part of photonic quantum technologies \cite{o2007optical,kimble2008quantum}, also spins with good spin-photon interfaces are promising candidates for storing information \cite{atature2018material}. These make color centers important for various quantum applications, including quantum communication, quantum sensing, and distributed quantum computing.

Ultra-bright and polarized single-photon emission from color centers in two-dimensional (2D) hexagonal boron nitride (\mbox{h-BN}) has been recently observed at room temperature \cite{tran2016quantum}. \mbox{h-BN} has attracted attention for several reasons. Firstly, it has a relatively large bandgap of around 6 eV  \cite{xia2014two,cassabois2016hexagonal,elias2019direct} which allows it to host many defects \cite{jungwirth2016temperature, exarhos2019magnetic,proscia2018near,konthasinghe2019rabi,tran2016robust}. However, the true atomic structure of most of these emitters remains unknown \cite{mendelson2021identifying,abdi2018color}.  Secondly, because of its 2D nature, it is promising for heterogeneous assembly and on-chip integration into devices \cite{grosso2017tunable, stern2021room}. Thirdly, some defects in \mbox{h-BN} might have high sensitivity to the environment because of their location at the surface, which is advantageous for quantum sensing applications \cite{reserbat2021quantum}. Finally, defects in \mbox{h-BN} are the only known solid-state sources that can display Fourier transform limited lines at room temperature \cite{dietrich2020solid}. If the Fourier transform of an emitter's temporal profile matches its spectral lineshape, then the emitter resonance does not fluctuate during the timescale of emission. This implies that quantum coherence is maintained so that the emitter can be used for many quantum protocols.

It has been shown that visible range SPEs in \mbox{h-BN} originate from carbon-related defects \cite{mendelson2021identifying}. Jara \textit{et al}. \cite{jara2021first} suggest that the neutral C\textsubscript{2}C\textsubscript{N} and C\textsubscript{2}C\textsubscript{B} carbon trimer defects might have zero-phonon line (ZPL) energies of 1.62 eV and 1.65 eV, respectively, and a phonon sideband of around 160 meV, which is typically found in many experiments \cite{mendelson2021identifying,hoese2020mechanical}. However, a new study suggests that the C\textsubscript{2}C\textsubscript{B} defect might have a ZPL energy of 1.36 eV \cite{auburger2021towards}. This energy is too far from the visible range, and so we focus only on the C\textsubscript{2}C\textsubscript{N} defect where both studies agree on a ZPL energy of around 1.6 eV.

In this study, we explore the electronic structure of the C\textsubscript{2}C\textsubscript{N} defect in 2D h-BN and find the possible radiative and non-radiative transitions to model the observed lines. To do so, we combine group theory analysis with density functional theory (DFT) calculations \cite{hepp2014electronic, sajid2018defect}. We determine the symmetry-adapted molecular orbitals (MO) using group theory analysis. Then, we use DFT results to determine the relative energy ordering of these orbitals \cite{doherty2011negatively}. Next, we obtain the total orbital and spin multi-electron states by filling the lowest energy MOs, which gives us the ground state. Exciting electrons to the higher energy MOs gives us the excited states \cite{maze2011properties}. We calculate the total energy of the electronic structures with DFT, and the difference between these energies gives us the transition energies between defect states.

We then consider the spin-orbit, the spin-spin, and external magnetic field interactions and find matrix elements of the Hamiltonian, where group theory decreases the complexity by reducing the number of non-zero elements. Furthermore, we look at the interaction between the defect and the electromagnetic field and find non-vanishing matrix elements to derive the optical transitions. Combining this with the spin-orbit and the spin-spin Hamiltonians gives us possible non-radiative transitions assisted by the spin-orbit and the spin-spin interactions \cite{abdi2018color}. We also examine the hyperfine interaction of the ground state with a possible nearby nuclear spin \cite{doherty2012theory}. Finally, we look at the dynamics of this system and simulate the optically detected magnetic resonance (ODMR) signal predicted by the Lindblad master equation \cite{manzano2020short}.

This paper is organised as follows. In Sec. \ref{sec:MO} we discuss the symmetry of the C\textsubscript{2}C\textsubscript{N} defect and determine the symmetry-adapted MOs. Then we investigate multi-electron states (Sec. \ref{sec:multielectron}), the spin-orbit interaction (Sec. \ref{sec:SO}), the spin-spin interaction (Sec. \ref{sec:ss}), spin-orbit and spin-spin mediated transitions (Sec. \ref{sec:SO-SS}), selection rules for the transitions (Sec. \ref{sec:selection_rules}), external magnetic field effect (Sec. \ref{sec:magentic}), and hyperfine interaction (Sec. \ref{sec:hyperfine}). In Sec. \ref{sec:ODMR} we simulate the ODMR spectra and the $g^2(\tau)$ second-order correlation function. Next, we provide a summary in Sec. \ref{sec:conclusion}. Finally, we discuss computational methods in Sec. \ref{sec:comp_detail}. Matrix elements of all of the interactions and more configurations for the ODMR simulations are given in the Supplementary Material \ref{sec:intro}. 

\section{Molecular orbitals \label{sec:MO}}
\begin{figure}
    \centering
    \subfigure[]{\label{fig:C_Tri}\includegraphics[scale=0.25]{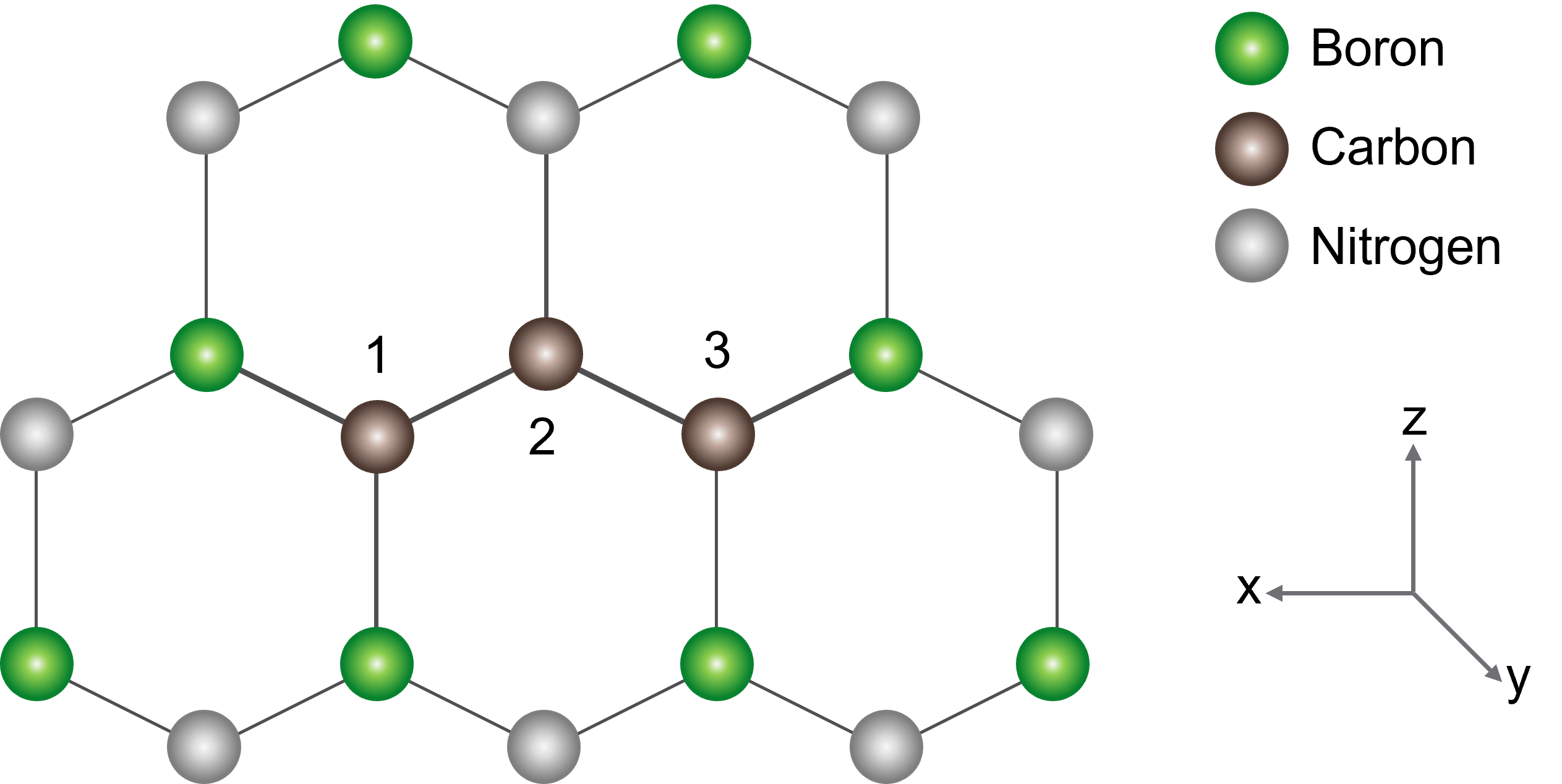}}
    \subfigure[]{\label{fig:C_2v}\includegraphics[scale=0.22]{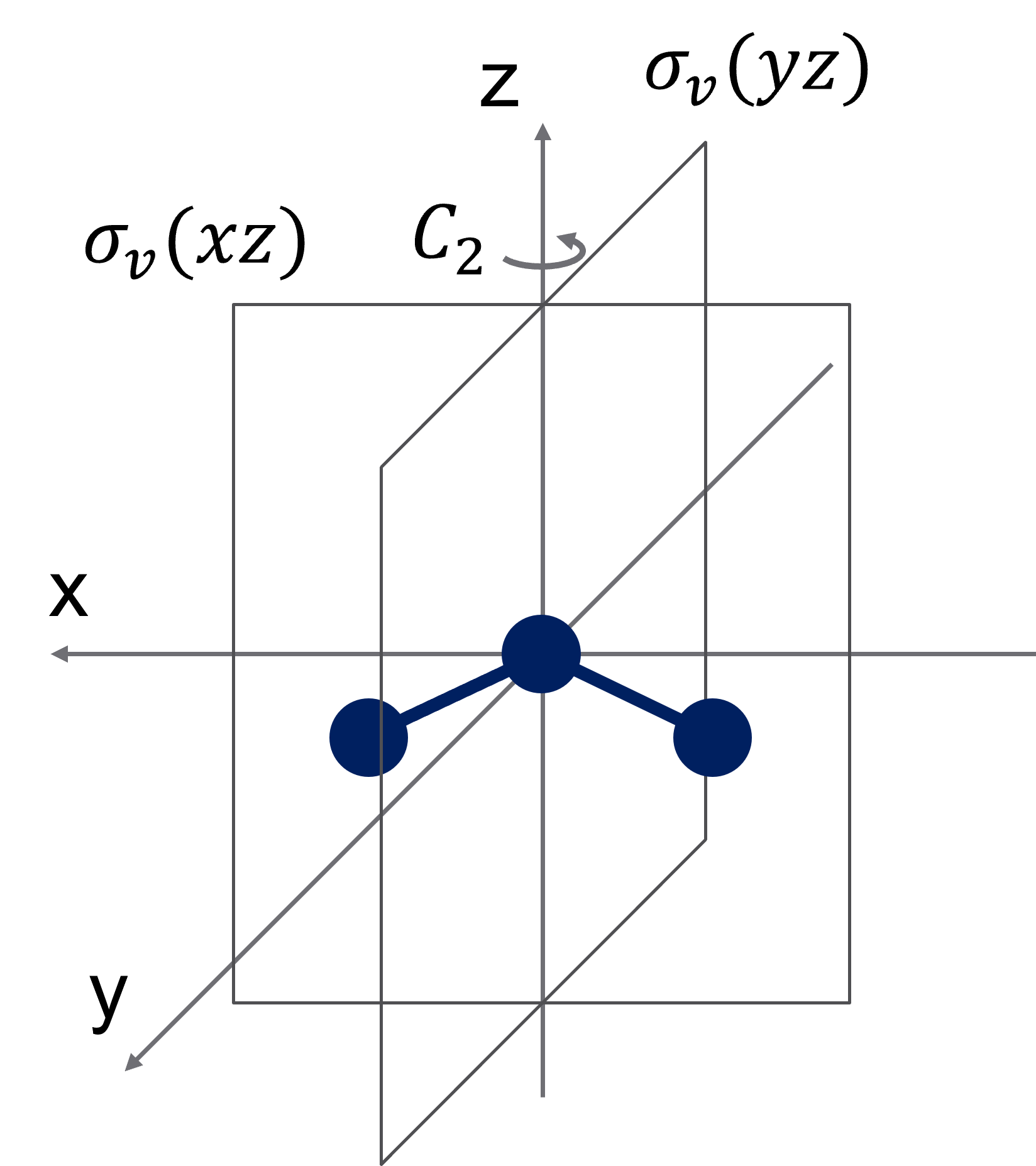}}
    \caption{(a) Symmetry operators of $C_{2v}$ point group, apart from the identity operator (E), shown for a carbon trimer defect. The first one is $C_{2}(z)$ which is a rotation by $\pi$ around z axis. The other two are reflections through xz and yz planes, respectively $\sigma_{v}(xz)$ and $\sigma_{v}(yz)$. Note that the three carbon atoms are in the $xz$ plane. (b) The atomic configuration of the C\textsubscript{2}C\textsubscript{N} defect in 2D \mbox{h-BN} sheet.}
\end{figure}

The atomic configuration of the C\textsubscript{2}C\textsubscript{N} defect is shown in Fig. \ref{fig:C_Tri}, where C\textsubscript{2} denotes the C\textsubscript{B}C\textsubscript{N} carbon dimer, and C\textsubscript{N} denotes a substitution of a nitrogen atom with a carbon atom. In order to find the symmetry group of the defect, it is important to know if the defect is in- or out-of-plane, as some defects might be distorted out of the plane \cite{noh2018stark}. A recent study suggests that distortion from the plane for the C\textsubscript{2}C\textsubscript{N} defect is negligible and that it has a planar structure \cite{auburger2021towards}. Thus, this defect has $C_{2v}$ symmetry, which is supported by defect wave functions as in Fig. \ref{fig:orbitals_a}.

\begin{table}[b]
    \centering\renewcommand{\arraystretch}{1.4}
    \begin{tabular}{c | c c c c || c c c}
        \hline
        $C_{2v}$ & $E$ & $C_2(z)$ & $\sigma _v(xz)$ & $\sigma _v(yz)$ & Linear & Quadratic & Cubic\\
        \hline
        $A_1$ & 1 & 1 & 1 & 1 & $z$ & $x^2,y^2,z^2$ & $z^3, x^2z, y^2z$\\
        $B_2$ & 1 & -1 & -1 & 1 & $y, R_x$ & $yz$ & $yz^2, y^3, x^2y$\\ 
        $B_1$ & 1 & -1 & 1 & -1 & $x, R_y$ & $xz$ & $xz^2, x^3, xy^2$\\
        $A_2$ & 1 & 1 & -1 & -1 & $R_z$ & $xy$ & $xyz$\\
        \hline
    \end{tabular}
    \caption{Character table for $C_{2v}$ point group. $E$, $C_2(z)$, $\sigma _v(xz)$, $\sigma _v(yz)$ are symmetry operators. $A_1$, $B_2$, $B_1$, and $A_2$ are IRs of the point group.}
    \label{tab:charc2v}
\end{table}

The ground-state configuration of carbon is $1s^2 2s^2 2p^2$. The planarity of the defect implies that carbon atoms will have $sp^2$ hybridization. In $sp^2$ hybridization, the $2s$ orbital is mixed with only two of the three available $2p$ orbitals. The third $2p$ orbital remains unhybridized and out of the plane and in the $\hat{y}$ direction, which is also confirmed by our DFT calculations shown in Fig. \ref{fig:orbitals_a}.

Each carbon atom of the C\textsubscript{2}C\textsubscript{N} defect shares three of its valence electrons with nearby atoms in the lattice; therefore, they each have one unpaired electron. Thus, the dangling bonds of the defect are $\pi$ bonds, and they are denoted by $\{\pi_1, \pi_2, \pi_3\}$.

Now, we need to find the symmetry-adapted MOs of this defect. The MOs are eigenfunctions of the Coulombic Hamiltonian.
We apply the projection operator,
\begin{equation}
    \label{eq:poperator}
    \phi_r=P^{(r)}\sigma_i=\frac{l_r}{h}\sum_{e}\chi_e^{(r)}R_e\pi_i,
\end{equation}
with a specific irreducible representation (IR) on our dangling bonds to find symmetrized MOs \cite{tinkham2003group}. Here, $P^{(r)}$ is the projection to the representation $r$, $l_r$ is the dimension of the representation $r$, $h$ is the number of symmetry group members, $\chi_e^{(r)}$ is the character of the operator $e$ in the representation $r$, $R_e$ is the symmetry operator, and $\pi_i$ is the dangling bond $i$. 
According to the character table of $C_{2v}$ point group (Table \ref{tab:charc2v}), $b$ and $b'$ MOs transform according to IR $B_2$. They are defined as
\begin{gather}
    b=\alpha\pi_2+\frac{\beta}{\sqrt{2}}(\pi_1+\pi_3),\\
    b'=\beta\pi_2+\frac{\alpha}{\sqrt{2}}(\pi_1+\pi_3),
\end{gather}
where $\alpha$ and $\beta$ are overlap integrals and     $|\alpha|^2+|\beta|^2=1$. There is another MO that transforms as IR $A_2$, defined as
\begin{gather}
    a=\frac{1}{\sqrt{2}}\{\pi_1-\pi_3\}.
\end{gather}

\section{Multi-electron states \label{sec:multielectron}}

\begin{figure*}[htp]
	\centering
	\subfigure[]{
	    \label{fig:orbitals_a}
	    \includegraphics[width=5.5 cm]{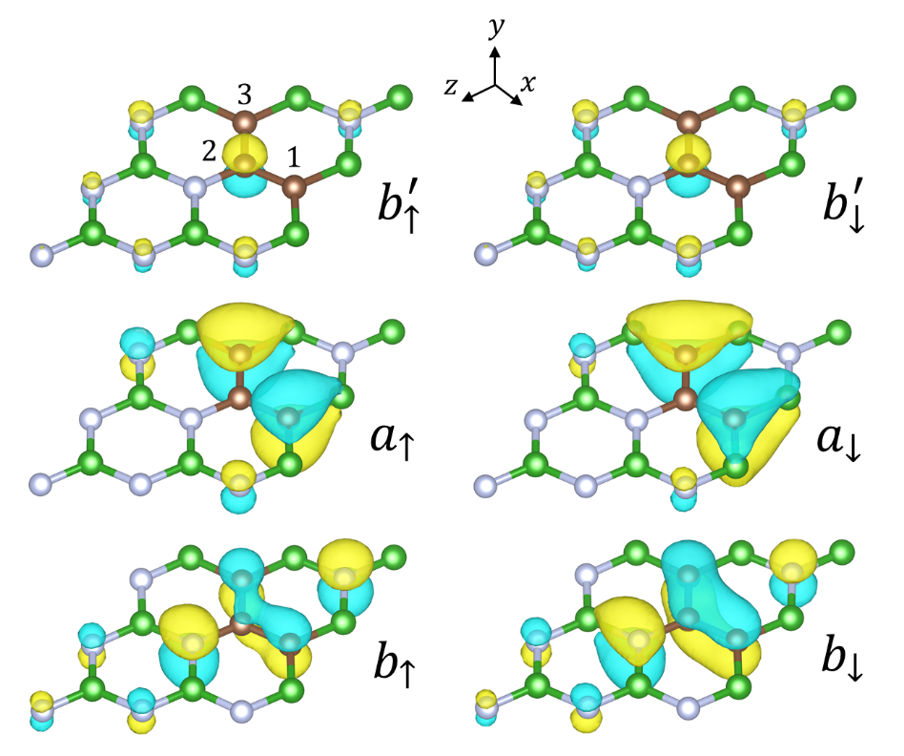}}
	\subfigure[]{
	    \label{fig:orbitals_b}
	    \includegraphics[width=11.5 cm]{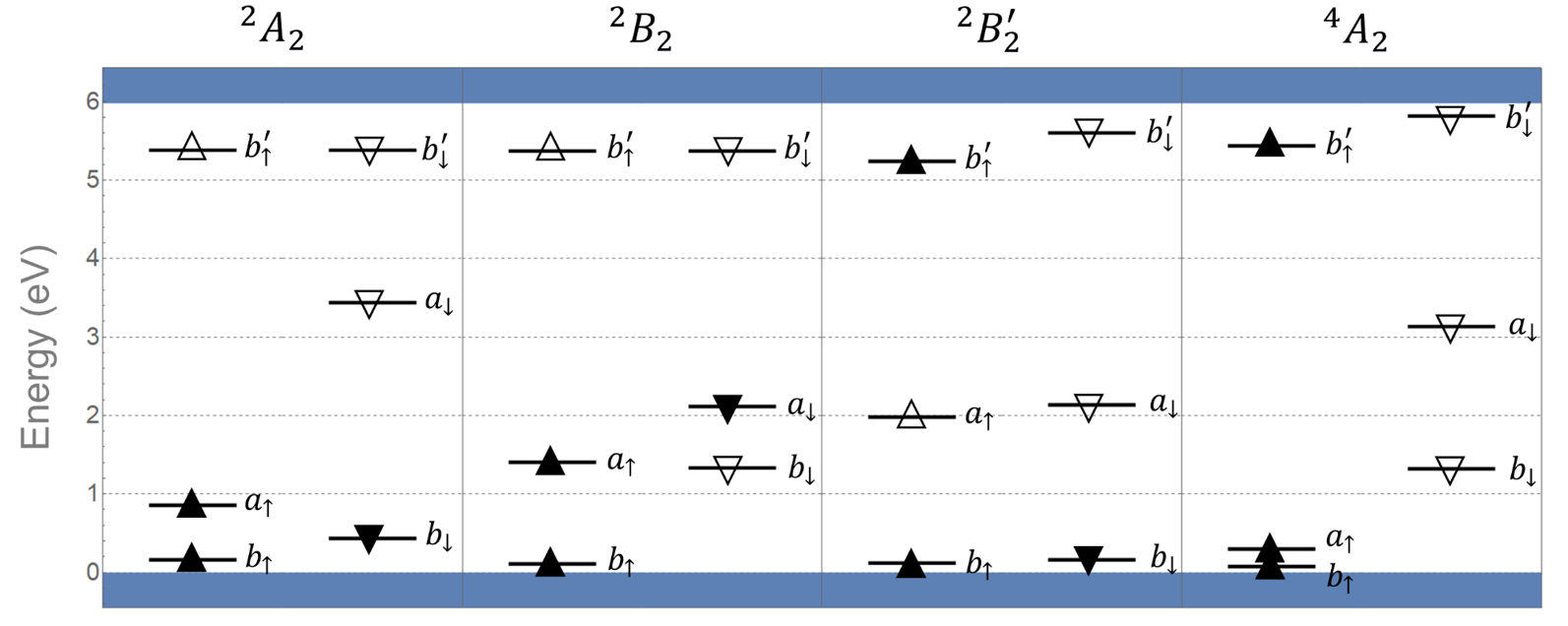}}
	\caption{(a) Ground-state wave functions of the C\textsubscript{2}C\textsubscript{N} defect. The positive (negative) components of each wave function are visualized by the yellow (blue) lobes. The corresponding symmetries are best represented when the $b$ and $a$ orbitals are plotted here at an isosurface level of $\pm 0.007 \text{\AA}^{-3}$, and the $b'$ orbital at $\pm 0.0002 \text{\AA}^{-3}$. The orbital energies increase from the bottom to the top, i.e., $E_b < E_a < E_{b'}$. Only the atoms and contributions to the wave function which are close to the C\textsubscript{2}C\textsubscript{N} defect are shown for simplicity. The carbon atoms are brown, boron atoms are green, and nitrogen atoms are grey. The diagrams were produced using VESTA \cite{momma2011vesta}. (b) Defect levels of the ground state and single-configuration excited states in the fundamental bandgap of \mbox{h-BN}. The occupied (unoccupied) levels are denoted by solid (empty) triangles.}
	\label{fig:orbitals}
\end{figure*}

We use DFT to find the energy of each of the MOs discussed above and their energy ordering. The defect wave functions in Fig. \ref{fig:orbitals_a} obtained from the DFT calculations show that the MO with the lowest energy transforms as IR $B_2$, so it represents the $b$ MO. This is because, according to the symmetry operators in Fig. \ref{fig:C_2v}, this MO is anti-symmetric under $C_2(z)$ and $\sigma_v(xz)$, and symmetric under $E$ and $\sigma_v(yz)$. The next MO with higher energy transforms as IR $A_2$, because it is anti-symmetric under $\sigma_v(xz)$ and $\sigma_v(yz)$, and symmetric under $E$ and $C_2(z)$. Therefore, it represents the $a$ MO. Finally, the one with the highest energy transforms as IR $B_2$ similar to the first one, and thus it represents the $b'$ MO.
Based on a previous study, the MOs in the ground state and the first excited state lie inside the bandgap \cite{jara2021first}.
Our \textit{ab initio} calculations show that the MOs in the next two excited states are also in the bandgap.

Multi-electron states are composed by filling the MOs with three unpaired electrons of the defect, starting from the lowest energy $b$ MO.
The $b$ MO will be fully occupied with two electrons in the ground state, and the $a$ MO will be half occupied. This configuration will form a spin doublet because the half occupied $a$ MO can be either spin up or down. So its spin multiplicity will be equal to 2. This lowest multi-electron state has the configuration $[b]^2[a]^1[b']^0$ which transforms as IR $A_2$. Other excited multi-electron states are produced by exciting each of these electrons to higher MOs. The $[b]^1[a]^2[b']^0$ and $[b]^2[a]^0[b']^1$ configurations are also spin doublets, similar to the ground state, and transform according to IR $B_2$. But the other excited state $[b]^1[a]^1[b']^1$ needs careful consideration. Since it is the addition of three spin 1/2 orbitals, it will have three irreducible spin representations, including one quartet state and two doublet states with multiplicities 4, 2, and 2, respectively. These states all transform as IR $A_2$. The corresponding electronic configurations of these states are given in Table \ref{tab:configuration} and the energy levels of the first four single-configuration states are given in Fig. \ref{fig:orbitals_b}.

\begin{table}[htp]
    \centering
    \begin{tabular}{c c c c}
        Configuration & $^{2S+1}\Gamma$ & Clebsch-Gordan states & Label\\
        \hline
        $[b]^2[a]^1[b']^0$ & $^2A_2$ & $|b\Bar{b}a\rangle, |b\Bar{b}\Bar{a}\rangle$ & $\mathcal{A}^{0,d}_{\pm1/2}$ \\
        $[b]^1[a]^2[b']^0$ & $^2B_2$ & $|ba\Bar{a}\rangle, |\Bar{b}a\Bar{a}\rangle$ & $\mathcal{B}^{1,d}_{\pm1/2}$ \\
        $[b]^2[a]^0[b']^1$ & $^2B'_2$ & $|b\Bar{b}b'\rangle, |b\Bar{b}\Bar{b}'\rangle$ & $\mathcal{B}^{2,d}_{\pm1/2}$ \\
        $[b]^1[a]^1[b']^1$ & $^4A_2$ & $|bab'\rangle, |\Bar{b}\Bar{a}\Bar{b}'\rangle$ & $\mathcal{A}^{3,q}_{\pm3/2}$ \\
         &  & $\frac{1}{\sqrt{3}}(|\Bar{b}ab'\rangle + |b\Bar{a}b'\rangle + |ba\Bar{b}'\rangle)$ & $\mathcal{A}^{3,q}_{+1/2}$ \\
         &  & $\frac{1}{\sqrt{3}}(|b\Bar{a}\Bar{b}'\rangle + |\Bar{b}a\Bar{b}'\rangle + |\Bar{b}\Bar{a}b'\rangle)$ & $\mathcal{A}^{3,q}_{-1/2}$ \\
         & $^2A_2'$ & $\frac{1}{\sqrt{6}}(|\Bar{b}ab' \rangle + |b\Bar{a}b'\rangle -2 |ba\Bar{b}' \rangle)$ & $\mathcal{A}^{3,d'}_{+1/2}$ \\
         &  & $\frac{1}{\sqrt{6}}(|\Bar{b}a\Bar{b}'\rangle + |\Bar{b}\Bar{a}b'\rangle -2 |b\Bar{a}\Bar{b}'\rangle)$ & $\mathcal{A}^{3,d'}_{-1/2}$ \\
         & $^2A_2''$ & $\frac{1}{\sqrt{2}}(|b\Bar{a}b' \rangle - |\Bar{b}ab'\rangle)$ & $\mathcal{A}^{3,d''}_{+1/2}$ \\
         &  & $\frac{1}{\sqrt{2}}(|b\Bar{a}\Bar{b}'\rangle - |\Bar{b}a\Bar{b}'\rangle)$ & $\mathcal{A}^{3,d''}_{-1/2}$
    \end{tabular}
    \caption{Configuration of total wave functions. Some of these states are entangled states which need careful consideration when calculating their energy using DFT. Spin-down electrons in an orbital are shown with a line over them. In the label column, calligraphic letters $\mathcal{A}$ and $\mathcal{B}$ represent IRs $A_2$ and $B_2$, respectively. Also, $d$ and $q$ in the superscript stand for doublet and quartet states, respectively. Prime and double prime in IRs of each state is used just to distinguish them with other states with the same IR.}
    \label{tab:configuration}
\end{table}

\section{Spin-orbit interaction \label{sec:SO}}
The spin-orbit interaction is the sum of the Larmor and Thomas interaction energy which is given by \cite{maze2011properties}
\begin{equation}
\begin{split}
    H_\text{SO}&=\sum_{k}\frac{\hbar}{2 m_\text{e}^2 c^2}(\nabla_k V\times \boldsymbol{p}_k)\cdot\left(\frac{\boldsymbol{s}_k}{\hbar}\right)\\
    &=\sum_{k}l_k\cdot\left(\frac{\boldsymbol{s}_k}{\hbar}\right),
\end{split}
\end{equation}
where $V$ is the electric potential energy of the nucleus, $\hbar$ is the reduced Planck constant, $m_e$ is the electron rest mass, $c$ is the speed of light in vacuum, $\boldsymbol{s}_k$ is the spin of electron $k$, $\boldsymbol{p}_k$ is the momentum of electron $k$, and $k$ sums over all electrons.
By utilizing group theory, we omit the vanishing components of the matrix elements of $l_k$. The elements $\langle\phi_i|l_k|\phi_j\rangle$ are non-vanishing only if $\Gamma(\phi_i)\otimes\Gamma(l_k)\otimes\Gamma(\phi_j)\supset\Gamma^{A_1}$, where $\Gamma$ is the irreducible representation. Since $\boldsymbol{l}$ is proportional to $\boldsymbol{r}\times\boldsymbol{p}$, it transforms as $(B_2, B_1, A_2)$. Based on Table \ref{tab:mat_el2}, only $l_y$, which transforms as IR $B_1$, will have non-zero values. Therefore,
\begin{gather}
    H_\text{SO}=\sum_{k}l_k^{(y)}\left(\frac{s_k^{(y)}}{\hbar}\right).
\end{gather}


Because of the symmetry of the system and according to Table \ref{tab:mat_el2}, we know that only elements in the form of $\langle B_2|H_{\text{so}}|A_2\rangle$ and their complex conjugate will be non-zero. Also, since we know $s_y=\frac{1}{2i}(s_+-s_-)$, only the states whose spin are different by one will yield non-zero values. After considering these symmetry constraints, we obtain the matrix elements provided in the appendix (Sec. \ref{sec:sup_HSO}).

\begin{table}[htp]
    \centering
    \begin{minipage}{.24\linewidth}
        \begin{tabular}{c | c c}
            $O^{A_1}$ & $B_2$ & $A_2$\\
            \hline
            $B_2$ & $\times$ & 0 \\
            $A_2$ & 0 & $\times$ \\ 
            \hline
        \end{tabular}
    \end{minipage}
        \begin{minipage}{.24\linewidth}
        \begin{tabular}{c | c c}
            $O^{B_2}$ & $B_2$ & $A_2$\\
            \hline
            $B_2$ & 0 & 0 \\
            $A_2$ & 0 & 0 \\ 
            \hline
        \end{tabular}
    \end{minipage}
        \begin{minipage}{.24\linewidth}
        \begin{tabular}{c | c c}
            $O^{B_1}$ & $B_2$ & $A_2$\\
            \hline
            $B_2$ & 0 & $\times$ \\
            $A_2$ & $\times$ & 0 \\ 
            \hline
        \end{tabular}
    \end{minipage}
        \begin{minipage}{.24\linewidth}
        \begin{tabular}{c | c c}
            $O^{A_2}$ & $B_2$ & $A_2$\\
            \hline
            $B_2$ & 0 & 0 \\
            $A_2$ & 0 & 0 \\ 
            \hline
        \end{tabular}
    \end{minipage}
    \caption{Matrix elements of operators with specific symmetries in the $\{B_2,A_2\}$ manifold where $\times$ indicates a non-zero value.}
    \label{tab:mat_el2}
\end{table}

\section{Spin-spin interaction \label{sec:ss}}
The spin-spin interaction is described by \cite{Reviews_Computational_Chemistry}

\begin{equation}
\begin{split}
    H_\text{ss}&=\frac{\mu_0\gamma_\text{e}^2\hbar^2}{4\pi}\sum_{i>j}\frac{1}{r_{ij}^3}[\boldsymbol{s}_i\cdot\boldsymbol{s}_j-3(\boldsymbol{s}_i\cdot\hat{\boldsymbol{r}}_{ij})(\boldsymbol{s}_j\cdot\hat{\boldsymbol{r}}_{ij})]\\
    &=\frac{\mu_0\gamma_\text{e}^2\hbar^2}{4\pi}\sum_{i>j}[\boldsymbol{s}_i\cdot\hat{D}_{ij}\cdot\boldsymbol{s}_j]\\
    &= \frac{\mu_0\gamma_\text{e}^2\hbar^2}{4\pi}\sum_{i>j}[\hat{\boldsymbol{s}}_{ij}^{(2)}\otimes\hat{D}_{ij}^{(2)}]^{(0)},
    \label{eq:Hss3}
\end{split}
\end{equation}
where $r_{ij}=r_i-r_j$ is the distance between electrons $i$ and $j$, $\hat{\boldsymbol{r}}_{ij}$ is the unit vector from electron $i$ to electron $j$, $\boldsymbol{s}_i$ is the spin of nucleus $i$, $\mu_0$ is the vacuum permeability, and $\gamma_\text{e}$ is the electron gyromagnetic ratio. $\hat{\boldsymbol{s}}_{ij}^{(2)}=\hat{\boldsymbol{s}}_{i}^{(1)}\otimes\hat{\boldsymbol{s}}_{j}^{(1)}$ is a rank two spin tensor and $\hat{D}_{ij}$ is a traceless second-rank tensor operator defined as,
\begin{gather}
    \hat{D}_{ij}=\frac{1}{r_{ij}^5}
    \begin{pmatrix}
        r_{ij}^2-3x_{ij}^2 & -3x_{ij}y_{ij} & -3x_{ij}z_{ij}\\ 
        -3x_{ij}y_{ij} & r_{ij}^2-3y_{ij}^2 & -3y_{ij}z_{ij}\\ 
        -3x_{ij}z_{ij} & -3y_{ij}z_{ij} & r_{ij}^2-3z_{ij}^2
    \end{pmatrix}.
    \label{eq:D2}
\end{gather}
Writing the interaction in this form simplifies the calculations of matrix elements.

For spherically symmetric states, traceless $\hat{D}_{ij}$ means all three diagonal elements vanish. However, due to the lack of spherical symmetry of this defect, we should consider these elements in this magnetic dipole-dipole interaction. More details and matrix elements of the spin-spin Hamiltonian are provided in the appendix Sec. \ref{sec:sup_HSS}.

\section{Spin-orbit and spin-spin induced transitions \label{sec:SO-SS}}
For the spin-orbit interaction, as we discussed before, only the matrix elements in the form of $\langle B_2|H_{\text{so}}|A_2\rangle$ and their complex conjugate will be non-zero. This indicates that there are no matrix elements in degenerate manifolds of $\{\mathcal{A}^{0,d},\mathcal{B}^{1,d},\mathcal{B}^{2,d},\mathcal{A}^{3,q}\}$. Therefore, there is no mixing due to the spin-orbit coupling. However, we have possible spin-orbit induced transitions between the states in these manifolds, which are $\mathcal{B}^{1,d} \leftrightarrow \mathcal{A}^{0,d}, \mathcal{B}^{2,d} \leftrightarrow \mathcal{A}^{0,d}, \mathcal{B}^{1,d} \leftrightarrow \mathcal{A}^{3,q}$, and $\mathcal{B}^{2,d} \leftrightarrow \mathcal{A}^{3,q}$. As discussed in Ref. \cite{goldman2015state,Phonon_induced_dynamic}, these types of transitions can happen in two steps. First, spin-orbit assisted transition occurs for example from $\mathcal{B}^{2,d}$ to a vibrational excited state of $\mathcal{A}^{3,q}$. This is followed by a relaxation to the vibrational ground-state, for example, via the emission of one or more phonons. Such a process will be possible if there is an overlap between the initial vibrational level of $\mathcal{B}^{2,d}$ and the excited vibrational level of $\mathcal{A}^{3,q}$.

Similarly and based on the findings of the previous section, the spin-spin interaction has no matrix element in the degenerate manifold of $\{\mathcal{A}^{0,d},\mathcal{B}^{1,d},\mathcal{B}^{2,d}\}$. However, there are non-zero matrix elements of the spin-spin interaction in the quartet state manifold. $\mathcal{D}_0$ is the diagonal, and $\mathcal{E}_3$ is the off-diagonal term. Hence, spin-spin interaction breaks the degenerate quartet states into two states and separates them by $2\mathcal{D}_0$. Also, the non-diagonal terms in the same manifold mix these two states. 
There are also possible spin-spin induced transitions between the states in these manifolds, which are $\mathcal{A}^{0,d} \leftrightarrow \mathcal{A}^{3,q}, \mathcal{B}^{1,d} \leftrightarrow \mathcal{A}^{3,q}, \mathcal{B}^{2,d} \leftrightarrow \mathcal{A}^{3,q}$.

\section{Selection rules \label{sec:selection_rules}}

\begin{figure*}
    \centering
    \subfigure[]{\label{fig:transitions}\includegraphics[width=8.5 cm]{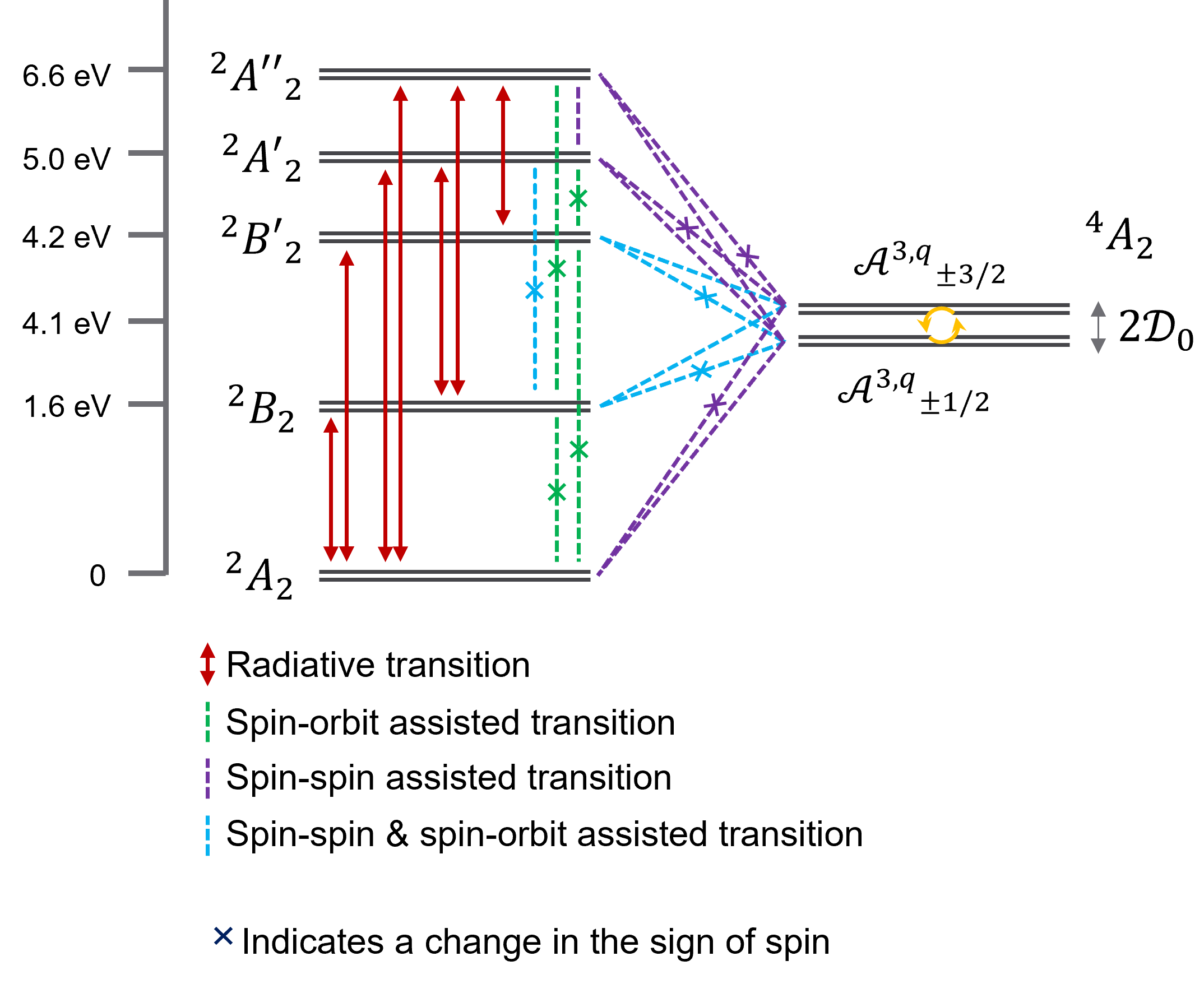}}
    \subfigure[]{\label{fig:anticrossing}\includegraphics[width=6.5cm]{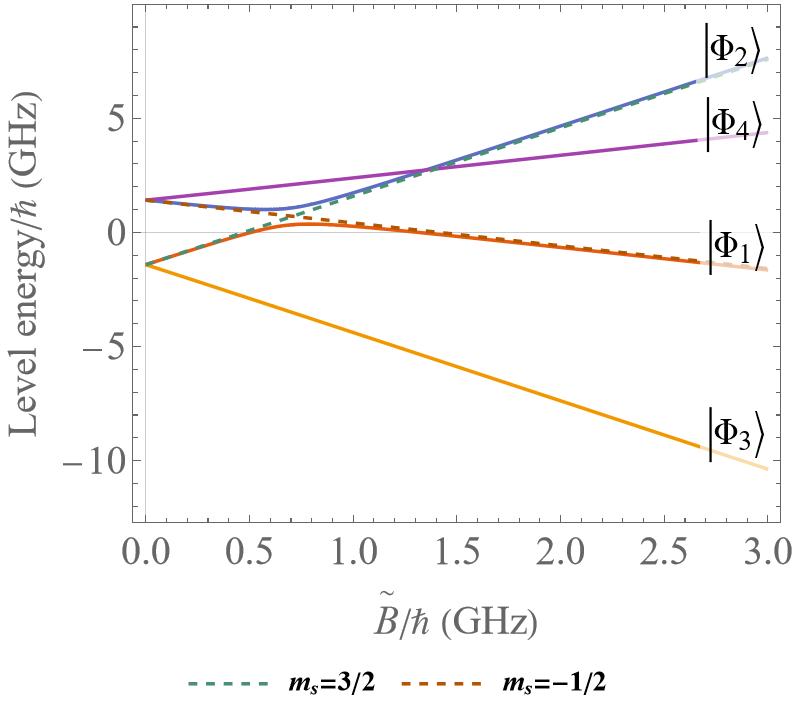}}
    \caption{(a) The electronic structure of the C\textsubscript{2}C\textsubscript{N} defect and possible radiative and non-radiative transitions. Red lines shows the possible electric dipole transitions. While dashed lines indicate possible phonon-assisted transitions. Yellow arrows show mixing between $A^{(3,q)}_{\pm1/2}$ and $A^{(3,q)}_{\pm3/2}$ due to the spin-spin coupling. The spin-spin coupling splits $^4A_2$ states by $2\mathcal{D}_0$ if we assume that $\mathcal{E}_3$ is much smaller than $\mathcal{D}_0$. The relative energy spacings of these states were obtained by our DFT calculations, which considers the Coulomb interaction and the HSE06 exchange-correlation functional. In this figure, we have assumed the quartet state is further detuned from the doublets than the spin-orbit coefficient. Usually the spin-orbit coefficient is on the order of GHz \cite{doherty2011negatively} and here the closest doublet to the quartet is separated by 0.1 eV corresponding to 24 THz. (b) The quartet state anticrosing, which shows an anticrossing between the $\ket{\Phi_1}$ and $\ket{\Phi_2}$ states near $\Tilde{B}_y/\hbar=0.7$ GHz. Here, we have assumed that $\Tilde{\mathcal{D}}_0$ and $\Tilde{\mathcal{E}}_3$ are equal to 1 GHz. The dashed lines show the behaviour of states with $m_s=3/2$ and $m_s=-1/2$ in the presence of a magnetic field.}
    \label{fig:results}
\end{figure*}

Here we look at the dominant transition allowed by the interaction of the electron with the electromagnetic field, which is the electric dipole transition. The electric dipole interaction is defined as
\begin{equation}
    H_\text{dipole} = \boldsymbol{E}\cdot\boldsymbol{d} = \sum_k e\boldsymbol{E}\cdot\boldsymbol{r}_k,
\end{equation}
where $\boldsymbol{E}$ is the electric field, $\boldsymbol{d}$ is the electric dipole moment, $\boldsymbol{r}_k$ is the position of $k$ electron with respect to the nucleus, and $e$ is the elementary electric charge. The position $\boldsymbol{r}$ in the $C_{2v}$ group transforms like $(B_1,B_2,A_1)$. Thus, according to Table \ref{tab:mat_el2}, the allowed transitions are induced by either $eE_xx$ or $eE_zz$ and the dipole moment lies completely in the plane. The dipole allowed transitions and the matrix elements are given in the appendix (Sec. \ref{sec:supp_dipole}). These results are summarized in Fig. \ref{fig:transitions}, which shows radiative and non-radiative transitions along with the energy levels of the states.

\section{External magnetic field \label{sec:magentic}}
In the presence of an external magnetic field, there will be another term for the Zeeman interaction of the magnetic field with spin and orbital angular momentum \cite{doherty2011negatively}. This interaction is given by
\begin{equation}
    H_\text{B}=\frac{e}{2m_\text{e}}\sum_k\left(\boldsymbol{l}_k+g_\text{e}\boldsymbol{s}_k\right)\cdot\boldsymbol{B},
\end{equation}
where $g_\text{e}$ is the electron spin g-factor, $\boldsymbol{s}$ is the electron spin, $\boldsymbol{l}$ is electron orbital angular momentum, $\boldsymbol{B}$ is the external magnetic field, and $k$ sums over all electrons.
But since $l_z$ transforms as IR $A_2$ and $l_x$ transforms as IR $B_2$, according to Table \ref{tab:mat_el2}, they do not contribute to the Hamiltonian. Therefore, the Zeeman interaction will be simplified to $H_\text{B}= \frac{e}{2m_\text{e}}\sum_k \left(B_xg_\text{e}s_{x,k} + B_y(l_{y,k}+g_\text{e}s_{y,k}) + B_zg_\text{e}s_{z,k}\right)$. The matrix elements of the Hamiltonian above are given in the appendix (Sec. \ref{sec:sup_HB}).

\subsection{Quartet state anticrossing}
As we discussed previously, the spin-spin interaction splits and mixes the quartet state eigenvalues and the spin-orbit interaction does not affect them. Adding a magnetic field perpendicular to the \mbox{h-BN} sheet ($\hat{y}$), modifies the energy eigenvalues of the quartet state. We add the matrix elements of the interactions for the quartet state from preceding sections and find its eigensystem. The energy eigenvalues are given by
\begin{eqnarray}
   &E_1=\Tilde{B}_y-\kappa_1,
   \\\nonumber
   &E_2=\Tilde{B}_y+\kappa_1,
   \\\nonumber
   &E_3=-\Tilde{B}_y-\kappa_1,
   \\\nonumber
   &E_4=-\Tilde{B}_y+\kappa_1,
\end{eqnarray}
and eigenvalues are given by
\begin{eqnarray}
   \ket{\Phi_1} &= \mu_1\ket{\mathcal{A}^{3,q}_{-1/2}}+i\mu_1\ket{\mathcal{A}^{3,q}_{+1/2}}
   \\\nonumber
   &\qquad\qquad
   +i\ket{\mathcal{A}^{3,q}_{-3/2}}+\ket{\mathcal{A}^{3,q}_{+3/2}},
   \\\nonumber
   \ket{\Phi_2} &= -\mu_2\ket{\mathcal{A}^{3,q}_{-1/2}}-i\mu_2\ket{\mathcal{A}^{3,q}_{+1/2}}
   \\\nonumber
   &\qquad\qquad
   +i\ket{\mathcal{A}^{3,q}_{-3/2}}+\ket{\mathcal{A}^{3,q}_{+3/2}},
   \\\nonumber
   \ket{\Phi_3} &= -\nu_1\ket{\mathcal{A}^{3,q}_{-1/2}}+i\nu_1\ket{\mathcal{A}^{3,q}_{+1/2}}
   \\\nonumber
   &\qquad\qquad
   -i\ket{\mathcal{A}^{3,q}_{-3/2}}+\ket{\mathcal{A}^{3,q}_{+3/2}},
   \\\nonumber
   \ket{\Phi_4} &= -\nu_2\ket{\mathcal{A}^{3,q}_{-1/2}}+i\nu_2\ket{\mathcal{A}^{3,q}_{+1/2}}
   \\\nonumber
   &\qquad\qquad
   -i\ket{\mathcal{A}^{3,q}_{-3/2}}+\ket{\mathcal{A}^{3,q}_{+3/2}},
\end{eqnarray}
where the coefficients are defined as
\begin{eqnarray}
    \kappa_1&=&\sqrt{4 \Tilde{B}_y^2+\Tilde{\mathcal{D}}_0^2+\Tilde{\mathcal{E}}_3^2-2 \Tilde{B}_y \left(\Tilde{\mathcal{D}}_0+\sqrt{3} \Tilde{\mathcal{E}}_3\right)},
    \\\nonumber
    \kappa_2&=&\sqrt{4 \Tilde{B}_y^2+\Tilde{\mathcal{D}}_0^2+\Tilde{\mathcal{E}}_3^2+2 \Tilde{B}_y \left(\Tilde{\mathcal{D}}_0+\sqrt{3} \Tilde{\mathcal{E}}_3\right)},
    \\\nonumber
    \mu_1 &=& \frac{\left(\sqrt{3} \Tilde{B}_y+\Tilde{\mathcal{E}}_3\right) \left(\kappa_1-\Tilde{B}_y+\Tilde{\mathcal{D}}_0\right)}{3 \Tilde{B}_y^2-\Tilde{\mathcal{E}}_3^2},
    \\\nonumber
    \mu_2 &=& \frac{\left(\sqrt{3} \Tilde{B}_y+\Tilde{\mathcal{E}}_3\right) \left(\kappa_1+\Tilde{B}_y-\Tilde{\mathcal{D}}_0\right)}{3 \Tilde{B}_y^2-\Tilde{\mathcal{E}}_3^2},
    \\\nonumber
    \nu_1 &=& \frac{\left(\sqrt{3} \Tilde{B}_y-\Tilde{\mathcal{E}}_3\right) \left(\kappa_2+\Tilde{B}_y+\Tilde{\mathcal{D}}_0\right)}{3 \Tilde{B}_y^2-\Tilde{\mathcal{E}}_3^2},
    \\\nonumber
    \nu_2 &=& \frac{\left(\Tilde{\mathcal{E}}_3-\sqrt{3} \Tilde{B}_y\right) \left(-\kappa_2+\Tilde{B}_y+\Tilde{\mathcal{D}}_0\right)}{\Tilde{\mathcal{E}}_3^2-3 \Tilde{B}_y^2}.
\end{eqnarray}
The variables with tilde are defined as below to simplify the equations.
\begin{eqnarray}
    \Tilde{B}_y&=&\frac{\gamma_\text{e}\hbar}{2}B_y
    \\\nonumber
    \Tilde{\mathcal{D}_0}&=&\frac{\mu_0\gamma_\text{e}^2\hbar^2}{16\pi}\mathcal{D}_0
    \\\nonumber
    \Tilde{\mathcal{E}_3}&=&\frac{\mu_0\gamma_\text{e}^2\hbar^2}{16\pi}\mathcal{E}_3
\end{eqnarray}

Based on these results and as shown in Fig. \ref{fig:anticrossing}, an anticrossing happens between $\ket{\Phi_1}$ and $\ket{\Phi_2}$ when the magnetic field compensates the spin-spin splitting at $\Tilde{B}_y$ near $\sqrt{\Tilde{\mathcal{D}}_0^2+\Tilde{\mathcal{E}}_3^2}\Big/2$. The $\ket{\Phi_3}$ state, remains unmixed as it is diverging from other states. The $\ket{\Phi_4}$ state is not mixed too, despite the fact that the $\ket{\Phi_2}$ state passes it at $\Tilde{B}_y$ near $\sqrt{\Tilde{\mathcal{D}}_0^2+\Tilde{\mathcal{E}}_3^2}$.

\section{Ground-state hyperfine interaction\label{sec:hyperfine}}
Nuclear spins in solids are a promising candidate for storing information and using them as quantum memories due to their long coherence time \cite{simon2010quantum}. Nuclear spin quantum memories have been demonstrated experimentally for the orbital ground state of the negatively-charged nitrogen-vacancy center in diamond \cite{fuchs2011quantum,shim2013room}. In this section, we will investigate the effect of the presence of a carbon-13 nuclear spin in the defect, which is given by $\hat{\Tilde{H}}=\hat{H}_\text{13C}+\hat{V}_\text{mhf} + \hat{V}_\text{ehf}$. The first term is Zeeman interaction of the nuclear spin with an external magnetic field, which is given by $\hat{H}_\text{13C} = - \gamma_\text{13C}\boldsymbol{B} \cdot \boldsymbol{\hat{I}}$, where $\boldsymbol{\hat{I}}$ is the nuclear spin and $\gamma_\text{13C}$ is the nuclear spin gyromagnetic ratio of $^{13}$C. The second (third) term is the electric (magnetic) component of the hyperfine interaction of the ground electronic state of the defect with the $^{13}$C nuclear spin \cite{auzinsh2019hyperfine,doherty2012theory}. We only have to look at the magnetic component since $^{13}$C has a nuclear spin of $I=1/2$, and the electric component is due to the quadrupole moment of nuclei with spin $I \geq 1$ \cite{stoneham2001theory}. Also, we ignored the nuclear spin-spin interactions in this paper. 

The magnetic hyperfine Hamiltonian accounts for the interaction between the nuclear spin and the electronic orbital magnetic moment in addition to the dipole-dipole interaction between the nuclear spin and the electron spin. The component of the hyperfine interaction that is related to the orbital angular momentum is given by $2g_I\mu_N\mu_B\frac{\mu_0\hbar}{4\pi}\sum_i\frac{1}{r_{i\text{C}}^3}\boldsymbol{I}\cdot \boldsymbol{L}$, where $\mu_N$ is the nuclear magneton, $\mu_B$ is the Bohr magneton, $g_I$ is the nuclear g-factor, and $r_{i\text{C}}$ is the distance between $^{13}$C and electron $i$. This component is  zero based on Table \ref{tab:mat_el2}, since our ground states transform as IR $B_2$ and do not have orbital angular momentum. Hence, we only need to consider the dipole-dipole interaction between the electron spin and the nuclear spin. The magnetic part of the hyperfine Hamiltonian, with these considerations, is given by
\begin{equation}
\begin{split}
    \hat{V}_\text{mhf}&=C_\text{mhf}\sum_i
    \Bigg\{\left(\frac{8\pi}{3}\delta(\hat{r}_{i\text{C}})-\frac{1}{r_{i\text{C}}^3}\right)\hat{s}_i\cdot\hat{I} +\\
    &\qquad\qquad
    \frac{3(\hat{s}_i\cdot\hat{r}_{i\text{C}})(\hat{I}\cdot\hat{r}_{i\text{C}})}{r_{i\text{C}}^5}\Bigg\}\\
        &=-C_\text{mhf}\sum_i\hat{s_i}\cdot \hat{A}_i^{(2)}\cdot \hat{I}\\
    &=-C_\text{mhf}\sum_i[\hat{J}_{i}^{(2)}\otimes\hat{A}_{i}^{(2)}]^{(0)},
    \label{eq:Vmhf1}
\end{split}
\end{equation}
where $C_\text{mhf}=g_I\mu_Ng_\text{e}\mu_B\frac{\mu_0\hbar^2}{4\pi}$, and $\hat{A}_i^{(2)}$ is a second rank tensor. The Fermi contact term contributes to the energy of orbitals with non-zero value of the wave function at the position of the nucleus. However, based on our DFT calculations (Fig. \ref{fig:orbitals_a}), the wave functions are zero at the position of the carbon nuclei and we can ignore the Dirac delta term. Consequently, the second order tensor $\hat{A}_i^{(2)}$ is given by 
\begin{gather}
    \hat{A}_{i}^{(2)}=\frac{1}{r_{i\text{C}}^5}
    \begin{pmatrix}
        r_{i\text{C}}^2-3x_{i\text{C}}^2 & -3x_{i\text{C}}y_{i\text{C}} & -3x_{i\text{C}}z_{i\text{C}}\\ 
        -3x_{i\text{C}}y_{i\text{C}} & r_{i\text{C}}^2-3y_{i\text{C}}^2 & -3y_{i\text{C}}z_{i\text{C}}\\ 
        -3x_{i\text{C}}z_{i\text{C}} & -3y_{i\text{C}}z_{i\text{C}} &  r_{i\text{C}}^2-3z_{i\text{C}}^2
    \end{pmatrix}.
    \label{eq:A2}
\end{gather}
For simplifying further calculations, we define $\hat{J}_{j}^{(2)}=\hat{s}_{i}\otimes\hat{I}$ and write the interaction in the compound tensor form.

According to Table \ref{tab:mat_el2}, for the ground states $\ket{\mathcal{A}^{0,d}_{\pm1/2}}$, only the operators of the form $O^{A_1}$ contributes to the hyperfine interaction. 
Thus, only the diagonal terms of $\hat{A}_{i}^{(2)}$ in Eq.~(\ref{eq:A2}) transform as IR $A_1$ contribute to the hyperfine interaction of the ground state, and the off-diagonal terms do not contribute. We write the basis of the ground state of the defect coupled to a $^{13}$C nuclear spin as

\begin{equation}
\begin{split}
    |\Psi^{g}_{1};1,+1\rangle&=\ket{\mathcal{A}^{0,d}_{+1/2}}|+\rangle_I,
    \\
    |\Psi^{g}_{2};1,0\rangle&=\frac{1}{\sqrt{2}}(\ket{\mathcal{A}^{0,d}_{+1/2}}|-\rangle_I+\ket{\mathcal{A}^{0,d}_{-1/2}}|+\rangle_I),
    \\
    |\Psi^{g}_{3};1,-1\rangle&=\ket{\mathcal{A}^{0,d}_{-1/2}}|-\rangle_I,
    \\
    |\Psi^{g}_{4};0,0\rangle&=\frac{1}{\sqrt{2}}(\ket{\mathcal{A}^{0,d}_{+1/2}}|-\rangle_I-\ket{\mathcal{A}^{0,d}_{-1/2}}|+\rangle_I).
\end{split}
\end{equation}

Based on the symmetry of the system, there can only be non-zero hyperfine matrix elements for states that have $\Delta S\in\{0,\pm2\}$. The results of the calculations for matrix elements are shown in the appendix (Sec. \ref{eq:Vmhf}). 


\section{ODMR signal \label{sec:ODMR}}

\begin{figure}[h]
    \centering
    \includegraphics[scale=0.65]{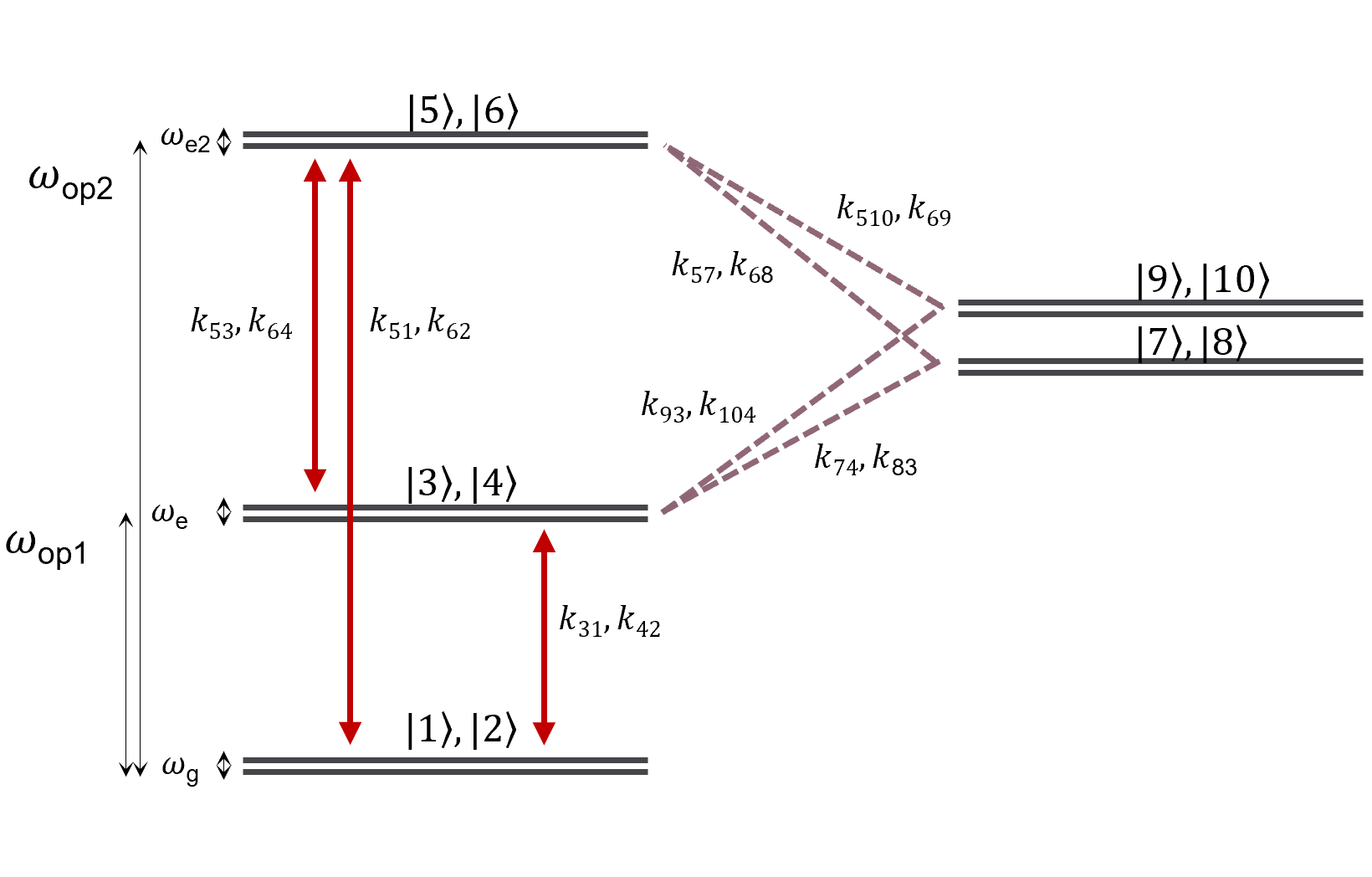}
    \caption{Our model for ODMR simulation. In this model, the energy spacing between levels, denoted by $\omega_{op1}$ and $\omega_{op2}$, are based on DFT calculations and rates, denoted by $k$, are approximated by rates from other studies on defects in \mbox{h-BN}. There are also spin splittings due to the external magnetic field, which are denoted by $\omega_{g}$, $\omega_{e}$, and $\omega_{e2}$. According to the matrix elements of the spin-orbit and spin-spin of this defect, the dashed lines are spin-dependent transitions which are important for observing an ODMR signal.}
    \label{fig:ODMR_Sim}
\end{figure}

There have been reports of ODMR signal for defects in \mbox{h-BN}. One of them is known to originate from the $V_B^-$ defect \cite{gottscholl2020initialization}, while the origins of the other observed ODMR signals are not established yet \cite{chejanovsky2021single, stern2021room}. Here, we present our results for the ODMR simulation using the model in Fig. \ref{fig:ODMR_Sim}. We used the Lindblad master equation to derive the second-order correlation function $g^{(2)}$ and the ODMR contrast.

We simulate a coherent pulse to excite the ground doublet state $\{\ket{1},\ket{2}\}$ to one of the two excited doublet states  $\{\ket{3},\ket{4}\}$ or $\{\ket{5},\ket{6}\}$. All of these electronic states are spin 1/2 states, and we assumed that their spin states are split due to an external magnetic field. We also simulate a coherent microwave pulse to probe the ODMR signal by changing its energy around the spin-splitting energy. 


If the quartet dark states $\{\ket{7},\ket{8},\ket{9},\ket{10}\}$ do not overlap with the phonon sideband of the first excited states $\{\ket{3},\ket{4}\}$, they can act as a meta-stable manifold during the decay of the second excited states $\{\ket{5},\ket{6}\}$. Since we predict that the quartet energy of around 4.1 eV is far above the 1.6 eV of the first excited doublet, it is unlikely that there is an overlap with the phonon sideband, which is typically smaller than about 500 meV for defects in \mbox{h-BN} \cite{mendelson2021identifying}. In addition, the decay to the quartet state and from the quartet state to the first excited state can depend on the spin, and hence the system could produce an ODMR signal. To see this signal, one should excite the ground-state levels to the second excited state doublet $\{\ket{5},\ket{6}\}$, after which a spin-independent non-radiative transition can occur into the quartet manifold.

The $^2B_2'$ state is the most promising candidate because its energy is very close to the quartet state, and most likely, it will overlap with its phonon sideband, allowing for fast non-radiative transitions to the meta-stable state. Thus, we restricted the model to the $^2B_2'$ level and ignored $^2A_2'$ and $^2A_2''$ levels because of their high energy. We also ignored non-radiative transitions related to $^2B_2' \leftrightarrow \ ^2A_2$, $^2B_2 \leftrightarrow ^2A_2$, and $^4A_2 \leftrightarrow ^2A_2$ because the energy difference between them is large and it is more likely that other non-radiative and radiative transitions will dominate the transitions between these states. In our simulation, the microwave Rabi frequency was chosen such that the ODMR signal had the highest value.

\begin{table}[hb]
    \centering
    \begin{tabular}{cc|cc}
        parameter & value & parameter & value\\
        \hline
        $\omega_\text{op2}$ & $\THztoeV{1015}$ &  $\omega_\text{op1}$ & $\THztoeV{387}$\\
        $\Omega_\text{op}$ & $5$ MHz & $\Omega_\text{MW}$ & $2$ MHz\\
        $k_{51}$ & $20$ MHz & $k_{62}$ & $20$ MHz\\
        $k_{31}$ & $1$ GHz  & $k_{42}$ & $1$ GHz \\
        $k_{53}$ & $10$ MHz & $k_{64}$ & $10$ MHz\\
        $k_{57}$ & $3$ MHz & $k_{68}$ & $1$ MHz\\
        $k_{74}$ & $0.01$ MHz & $k_{83}$ & $1$ MHz\\
        $k_{510}$ & $2.4\times10^{-7}$ MHz & $k_{69}$ & $2\times10^{-5}$ MHz\\
        $k_{93}$ & $2.1\times10^{-6}$ MHz & $k_{104}$ & $1.2\times10^{-5}$ MHz\\
        $\omega_\text{g}$ & $0.700$ GHz & $\omega_\text{e2}$ & $0.703$ GHz\\
        $\omega_\text{e}$ & $0.705$ GHz
    \end{tabular}
    \caption{Parameters used in the ODMR simulation. $\Omega_\text{op}$ is the coherent optical driving Rabi frequency, $\Omega_\text{MW}$ is the coherent microwave driving Rabi frequency, and other parameters are shown in Fig. \ref{fig:ODMR_Sim}.}
    \label{tab:ODMR_par}
\end{table}

\begin{figure}[h]
    \centering
    \includegraphics[width=8cm]{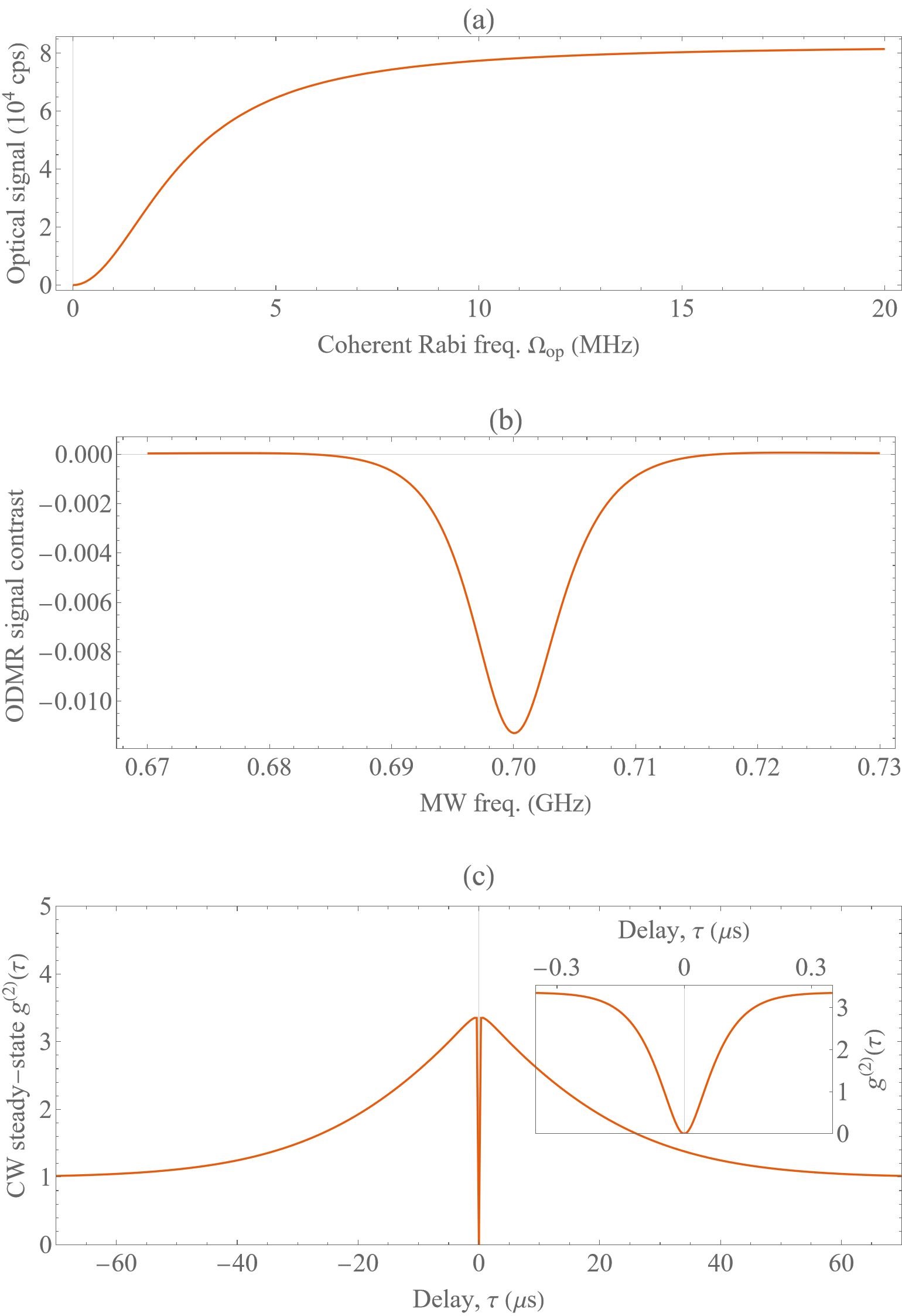}
    \caption{(a) Optical signal versus coherent Rabi frequency, which shows saturation near 20 MHz. The grid line shows the Rabi frequency that we used for the ODMR signal and the $g^2$ function. This is the frequency where the optical signal is near 75 \% of the saturation point. (b) ODMR signal, which shows 1.2 \% contrast at 700 MHz MW frequency. (c) The second-order correlation function, which shows significant bunching at microsecond timescales due to the meta-stable quartet state, and anti-bunching pattern at $\tau=0$. Inset: The same $g^2$ function for smaller timescales.}
    \label{fig:One_photon}
\end{figure}

The results obtained by exciting the ground-state levels directly to the second excited state $\{\ket{5},\ket{6}\}$, which is the $^2B'_2$ state, are given in Fig. \ref{fig:One_photon}. We have used the parameters in Table \ref{tab:ODMR_par} to predict this ODMR signal. The lifetime of excited states with energies around 2 eV has been measured for many emitters, and they are around a few ns \cite{jungwirth2017optical}, which we have used for the optical decay rates in our model. The intersystem crossing rates and meta-stable decay rates for which we see ODMR signal are on the order of magnitude of the rates seen in other defects \cite{boll2020photophysics}, but more calculations are needed to verify if the rates are in the proper range for the C\textsubscript{2}C\textsubscript{N} defect.

\begin{figure}[h]
    \centering
    \includegraphics[width=7cm]{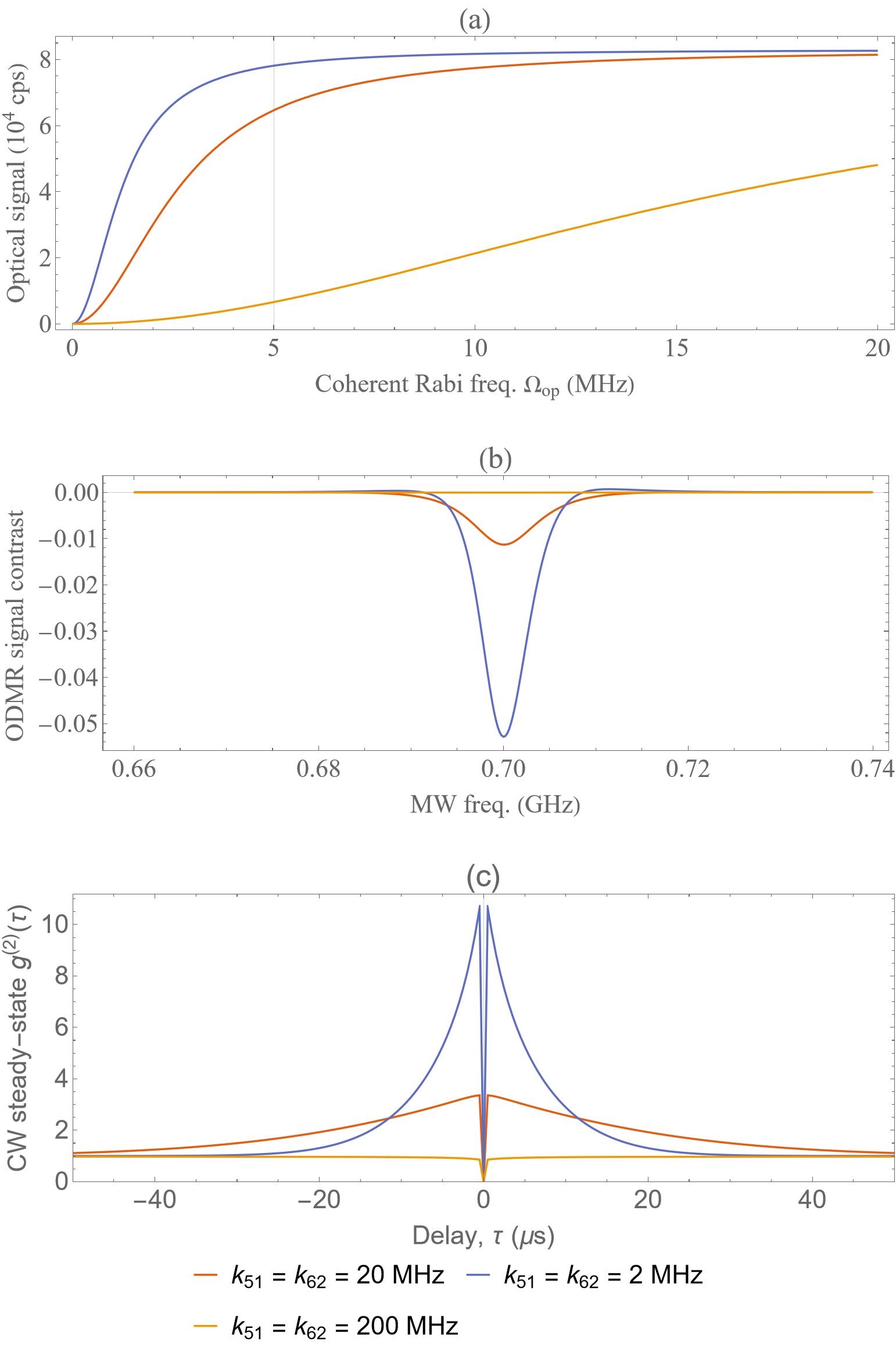}
    \caption{The effect of changing $k_{51}$ and $k_{62}$ on the optical signal, ODMR contrast, and the second-order correlation function. The grid line shows the Rabi frequency at 5 MHz. The results show that as $k_{51}$ and $k_{62}$ rates decrease, the magnitude of the ODMR signal increases.}
    \label{fig:k51=k62}
\end{figure}

\begin{figure}[h]
    \centering
    \includegraphics[width=7cm]{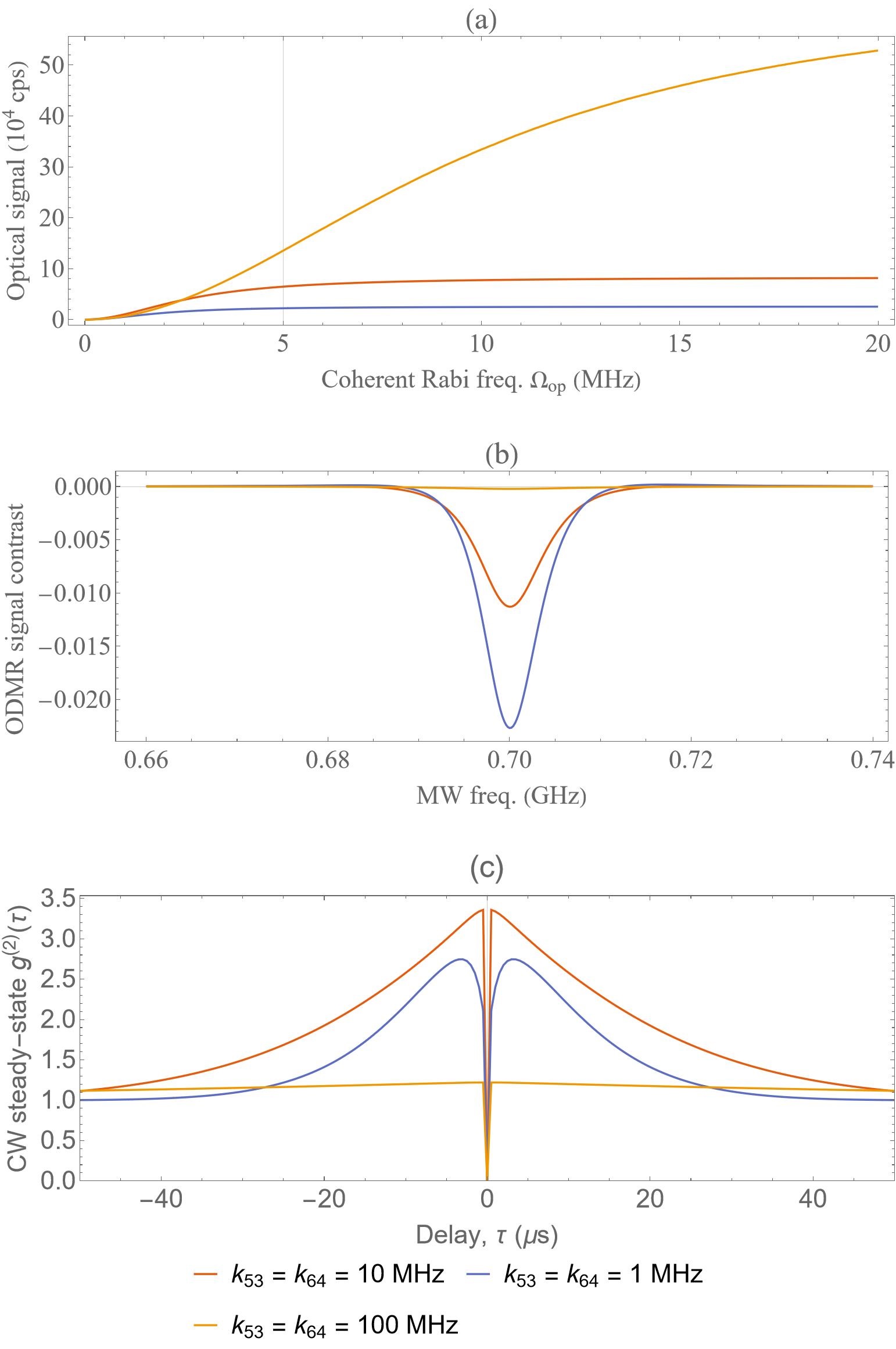}
    \caption{The effect of changing $k_{53}$ and $k_{64}$ on the optical signal, ODMR contrast, and the second-order correlation function. The grid line shows the Rabi frequency at 5 MHz. The results show that as $k_{53}$ and $k_{64}$ rates decrease, the magnitude of the ODMR signal increases.}
    \label{fig:k53=k64}
\end{figure}

\begin{figure}[h]
    \centering
    \includegraphics[width=7cm]{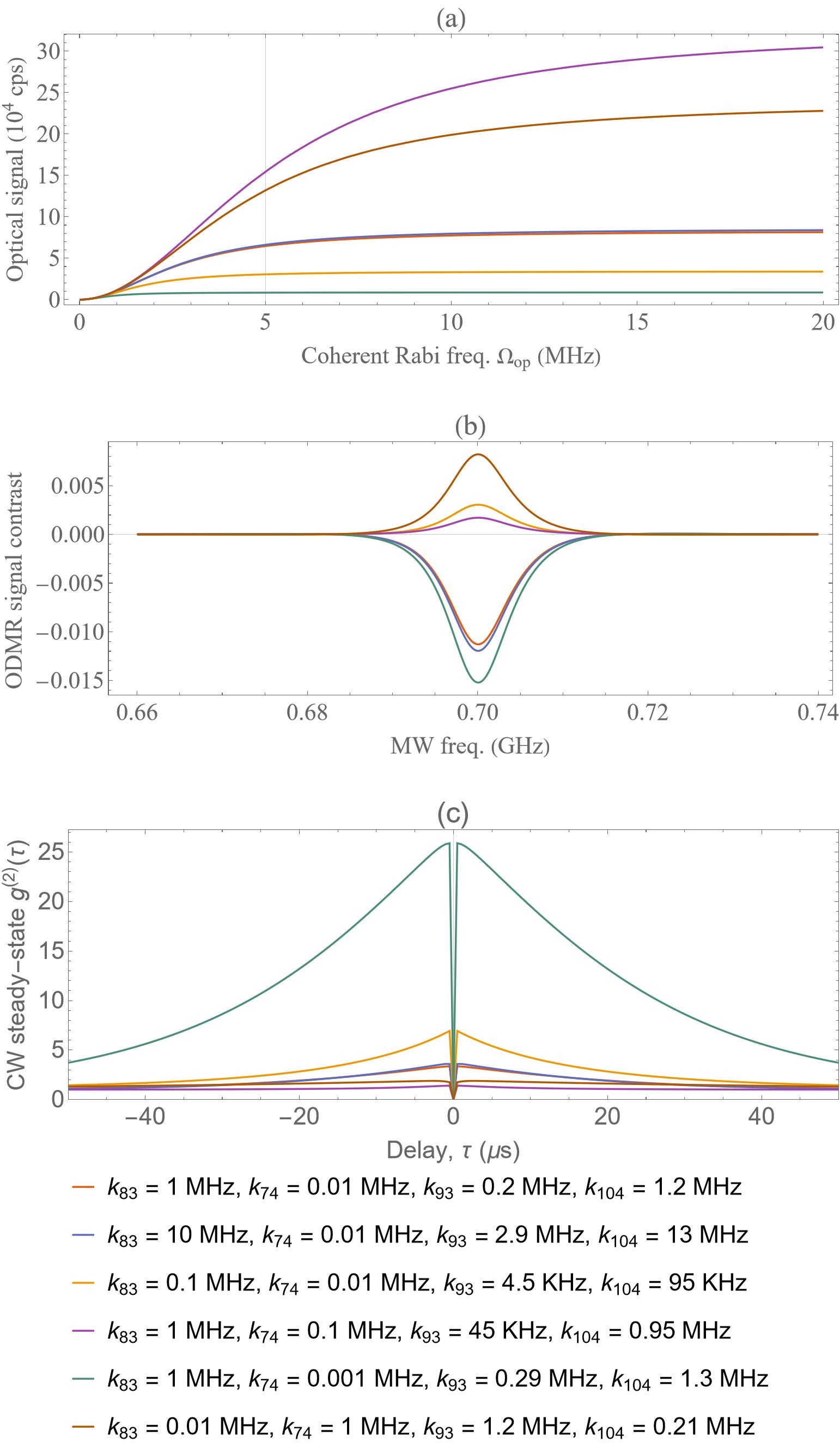}
    \caption{The effect of changing $k_{83}$ and $k_{74}$ on the optical signal, ODMR contrast, and the second-order correlation function. The grid line shows the Rabi frequency at 5 MHz. The results show that for some values of $k_{83}$ and $k_{74}$ rates, the sign of the ODMR contrast will change to positive.}
    \label{fig:k83_k74}
\end{figure}

Varying the parameters used for this model shows that $k_{51}$, $k_{62}$, $k_{53}$, and $k_{64}$ have major effects on the ODMR signal. They should be on the order of the values specified in Table \ref{tab:ODMR_par} in order to cause an ODMR signal contrast around 1\%. Their effect on the ODMR signal and $g^2$ correlation function is shown in Fig. \ref{fig:k51=k62} and Fig. \ref{fig:k53=k64}, respectively. Also, the difference between $k_{74}$ and $k_{83}$ is essential for having an ODMR signal. Based on our calculations for the spin-orbit and the spin-spin interactions, the matrix elements responsible for these transitions are different for the spin up and down. The transition amplitudes are proportional to these matrix elements, allowing the defect to have spin-dependent decay rates from and to the quartet state. 
The effect of changing $k_{74}$ and $k_{83}$ on the ODMR signal and $g^2$ correlation function is shown in Fig. \ref{fig:k83_k74}. Based on the matrix elements of the spin-orbit and spin-spin interactions, the $k_{74}$ and $k_{83}$ rates are related to the $k_{93}$ and $k_{104}$ rates. So changing each of them will affect the other two.

\section{Computational details\label{sec:comp_detail}}
The DFT calculations and post-processing were performed using the QUANTUM ESPRESSO open-source software package \cite{giannozzi2009quantum}. The calculations utilized a plane-wave basis set with a kinetic energy cutoff of 350 eV and projector augmented-wave pseudopotentials \cite{blochl1994projector}. All relaxation calculations were performed with a force convergence threshold of $10^{-4}$ eV/\AA.  Experimental investigations of point defects in \mbox{h-BN} typically consider multi-layer samples; however, it has been shown that DFT calculations result in negligible differences between the electronic structure of defects in single- and multi-layer systems \cite{tran2016quantum}. Our supercell consists of 98 atoms and a vacuum separation of 15 \AA\ between layers, corresponding to $7 \times 7$ unit cells of mono-layer \mbox{h-BN}. The atomic positions and in-plane lattice constant for the pristine \mbox{h-BN} structure were relaxed using the Perdew-Burke-Ernzerhof (PBE) exchange-correlation functional \cite{perdew1996generalized}. An in-plane lattice constant of $a = 2.5 $ \AA\ was obtained, consistent with previous findings \cite{ferreira2019excitons}. The Heyd-Scuseria-Ernzerhof (HSE06) hybrid functional \cite{heyd2003hybrid} was then used to optimize the direct bandgap at the $K$ high symmetry point \cite{wickramaratne2018monolayer} to the bulk value of $\approx 6$ eV \cite{abdi2018color}. A bandgap of 5.98 eV was obtained by setting the mixing parameter to 0.32 and fine-tuning the screening parameter to 
0.086 \AA$^{-1}$.

The C\textsubscript{2}C\textsubscript{N} defect was then added to the hexagonal lattice, the atomic positions were relaxed in-plane, and the ground-state wave functions of the single-particle defect levels were calculated.
Next, the single-configuration excited states ($^2A_2$, $^2B_2$, $^2B_2^\prime$, and $^4A_2$) were created using the $\Delta$SCF method \cite{gali2009theory}, and the atomic positions of each excited state electronic configuration were relaxed in-plane. The transition energies between defect states were calculated by considering the difference in total energies of the structures, obtained via spin-polarized calculations performed within the $\Gamma$-point approximation. The HSE06 functional has been shown to provide accurate results for defects in \mbox{h-BN} which exhibit low correlation and charge transfer, and as such, it is expected that the error in the DFT calculations of the single-configuration states is on the order of 0.1 eV \cite{reimers2018understanding, jara2021first}. The remaining states of interest ($^2A_2^\prime$ and $^2 A_2 ^{\prime\prime}$) are multi-configuration states which cannot be modeled in the DFT calculations using the $\Delta$SCF method. Rough estimates for the corresponding transitions energies were obtained following the method of Ref. \cite{ess2011singlet, mackoit2019carbon}, making use of the single-configuration states $\ket{b a b^\prime}$, $\ket{b a \overline{b^\prime}}$, and $\ket{b \overline{a} b^\prime}$ which were created within the $\Delta$SCF procedure (see appendix Sec. \ref{sec:singlet-triplet} for detailed calculations).

\section{Conclusion \label{sec:conclusion}}
We have used group theory and DFT calculations to find the electronic structure and transitions of the C\textsubscript{2}C\textsubscript{N} defect in 2D \mbox{h-BN}. The results are summarized in Fig. \ref{fig:transitions} which shows that there are several radiative transitions together with spin-orbit and spin-spin assisted non-radiative transitions. Also, the spin-spin interaction causes a splitting between quartet states $^4A_2$. We studied the effect of an external magnetic field and found that in the presence of an external magnetic field perpendicular to the plane, there is an anticrossing between the states of the quartet manifold. We also looked at the ground-state hyperfine interactions, which can be useful in future studies. Finally, we simulated the system using the Lindblad master equation. Although our results indicate that it is unlikely for the C\textsubscript{2}C\textsubscript{N} defect to be responsible for the ODMR signals that have been reported so far, we show that it could be possible to see an ODMR signal contrast of $\sim 1\%$ for the configuration discussed in the text. Properties of the defect that we considered are essential for future applications, e.g., for quantum networks and quantum sensing.\newline


\begin{acknowledgments}
We thank Igor Aharonovich for valuable discussions. This work was supported by the Natural Sciences and Engineering Research Council (NSERC) of Canada through its Discovery Grant and Strategic Project Grants Programs, the Alberta Major Innovation Fund Quantum Technology project,
Compute Canada, and Advanced Research Computing (ARC) IT team of the University of Calgary.
\end{acknowledgments}

\appendix
\section{Matrix elements of the Hamiltonian}
In this section, we provide the matrix elements of the interactions discussed in the main text.
\subsection{Spin-orbit interaction}\label{sec:sup_HSO}
The matrix elements of the spin-orbit interaction are given below, where the variables $\lambda$, and $\lambda'$ are defined as
\begin{equation}
    \lambda = \langle a|l^{(y)}|b\rangle, \lambda' = \: \langle b'|l^{(y)}|a\rangle.
\end{equation}

\begin{widetext}
\begin{gather}
    H_\text{so}=\frac{i}{2}\times
    \label{eq:HSO}\\
    \begin{blockarray}{ccccccccccccccc}
        \qquad\qquad\qquad & \mathcal{A}^{0,d}_{+1/2} & \mathcal{A}^{0,d}_{-1/2} & \mathcal{B}^{1,d}_{+1/2} & \mathcal{B}^{1,d}_{-1/2} & \mathcal{B}^{2,d}_{+1/2} & \mathcal{B}^{2,d}_{-1/2} & \mathcal{A}^{3,q}_{+1/2} & \mathcal{A}^{3,q}_{-1/2} & \mathcal{A}^{3,q}_{+3/2} & \mathcal{A}^{3,q}_{-3/2} & \mathcal{A}^{3,d'}_{+1/2} & \mathcal{A}^{3,d'}_{-1/2} & \mathcal{A}^{3,d''}_{+1/2} & \mathcal{A}^{3,d''}_{-1/2}\\
        \begin{block}{c(cccccccccccccc)}
            \mathcal{A}^{0,d}_{+1/2} & 
            0\\
            \mathcal{A}^{0,d}_{-1/2} & 
            0 & 0\\
            \mathcal{B}^{1,d}_{+1/2} & 
            0 & -\lambda & 0\\
            \mathcal{B}^{1,d}_{-1/2} & 
            \lambda & 0 & 0 & 0\\
            \mathcal{B}^{2,d}_{+1/2} & 
            0 & -\lambda' & 0 & 0 & 0\\
            \mathcal{B}^{2,d}_{-1/2} & 
            \lambda' & 0 & 0 & 0 & 0 & 0\\
            \mathcal{A}^{3,q}_{+1/2} & 
            0 & 0 & 0 & -\lambda'/\sqrt{3} & 0 & -\lambda/\sqrt{3} & 0\\
            \mathcal{A}^{3,q}_{-1/2} & 
            0 & 0 & -\lambda'/\sqrt{3} & 0 & -\lambda/\sqrt{3} & 0 & 0 & 0\\
            \mathcal{A}^{3,q}_{+3/2} & 
            0 & 0 & -\lambda' & 0 & -\lambda & 0 & 0 & 0 & 0\\
            \mathcal{A}^{3,q}_{-3/2} & 
            0 & 0 & 0 & -\lambda' & 0 & -\lambda & 0 & 0 & 0 & 0\\
            \mathcal{A}^{3,d'}_{+1/2} & 0 & 0 & 0 & -\lambda'/2\sqrt{3} & 0 & \lambda/\sqrt{3} & 0 & 0 & 0 & 0 & 0\\
            \mathcal{A}^{3,d'}_{-1/2} & 0 & 0 & -\lambda'/2\sqrt{3} & 0 & \lambda/\sqrt{3} & 0 & 0 & 0 & 0 & 0 & 0 & 0 \\
            \mathcal{A}^{3,d''}_{+1/2}  & 0 & 0 & 0 & \lambda'/2 & 0 & 0 & 0 & 0 & 0 & 0 & 0 & 0 & 0\\
            \mathcal{A}^{3,d''}_{-1/2} & 0 & 0 & \lambda'/2 & 0 & 0 & 0 & 0 & 0 & 0 & 0 & 0 & 0 & 0 & 0\\
        \end{block}
    \end{blockarray}
    \notag
\end{gather}
\end{widetext}


\subsection{Spin-spin interaction}\label{sec:sup_HSS}
We can see from Table \ref{tab:mat_el2} that the only elements of D that transform as IRs $A_1$ and $B_1$ would yield non-zero values. These are $\hat{D}^{A_1} = \{\hat{D}_{xx}, \hat{D}_{yy}, \hat{D}_{zz}\}$ and $\hat{D}^{B_1} = (\hat{D}_{xz} + \hat{D}_{zx})/2$.
In this interaction, we have the product of two rank-2 tensors. This product can be reduced to a sum of rank 4, 3, 2, 1, and 0 irreducible tensor operators. Nevertheless, here we need a compound tensor operator of rank zero. This tensor product is given by the equation below \cite{Reviews_Computational_Chemistry}.

\begin{equation}
    [\hat{\boldsymbol{s}}^{(2)}\otimes\hat{D}^{(2)}]^{(0)}_0 = \frac{1}{\sqrt{5}}\sum^{+2}_{q=-2}(-1)^{2-q}\hat{\boldsymbol{s}}_{-q}^{(2)}\hat{\boldsymbol{D}}_q^{(2)}
    \label{eq:zero_tensor}
\end{equation}

\begin{table}[htp]
    \centering
    \begin{tabular}{c c c}
        \hline
        Compound Tensor & q & Spherical Component\\
        \hline
        $\{\hat{s}^{(1)}_i\otimes\hat{s}_j^{(1)}\}^{(2)}_q$ & +2 &
        $\hat{s}^{(1)}_{i,+1}\hat{s}^{(1)}_{j,+1}$ \\
        & +1 & $(\hat{s}^{(1)}_{i,+1}\hat{s}^{(1)}_{j,0} + \hat{s}^{(1)}_{i,0}\hat{s}^{(1)}_{j,+1})/\sqrt{2}$\\
        & 0 & $\frac{1}{\sqrt{6}}(\hat{s}^{(1)}_{i,-1}\hat{s}^{(1)}_{j,+1} + 2\hat{s}^{(1)}_{i,0}\hat{s}^{(1)}_{j,0} + \hat{s}^{(1)}_{i,+1}\hat{s}^{(1)}_{j,-1})$\\
        & -1 & $(\hat{s}^{(1)}_{i,-1}\hat{s}^{(1)}_{j,0} + \hat{s}^{(1)}_{i,0}\hat{s}^{(1)}_{j,-1})/\sqrt{2}$\\
        & -2 & $\hat{s}^{(1)}_{i,-1}\hat{s}^{(1)}_{j,-1}$\\
        \hline
    \end{tabular}
    \caption{Second rank spin tensor.}
    \label{tab:compound_tensor}
\end{table}

In order to use Eq.~(\ref{eq:zero_tensor}), we need the spherical components of $\hat{\boldsymbol{D}}$. Spherical and Cartesian components of $\hat{\boldsymbol{D}}$ are related by the equations below.

\begin{equation}
    \begin{split}
    \hat{\boldsymbol{D}}^{(2)}_{\pm2} &= (\hat{\boldsymbol{D}}^{(2)}_{xx} - \hat{\boldsymbol{D}}^{(2)}_{yy} \pm 2i\hat{\boldsymbol{D}}^{(2)}_{xy})/\sqrt{2}\\
    \hat{\boldsymbol{D}}^{(2)}_{\pm1} &= \mp(\hat{\boldsymbol{D}}^{(2)}_{xz} \pm i\hat{\boldsymbol{D}}^{(2)}_{yz})\\ \hat{\boldsymbol{D}}^{(2)}_{0} &= (2\hat{\boldsymbol{D}}^{(2)}_{zz} - \hat{\boldsymbol{D}}^{(2)}_{xx} - \hat{\boldsymbol{D}}^{(2)}_{yy})/\sqrt{6}
    \end{split}
\end{equation}

The components of $\boldsymbol{\hat{s}}^{(1)}$ in the notation of spherical tensor operators are given by

\begin{eqnarray}
    \boldsymbol{\hat{s}}^{(1)}_{+1} &=& -\frac{1}{\sqrt{2}}(\boldsymbol{\hat{s}}_x+i\boldsymbol{\hat{s}}_y),
    \\\nonumber
    \boldsymbol{\hat{s}}^{(1)}_0&=&\boldsymbol{\hat{s}}_z,
    \\\nonumber
    \boldsymbol{\hat{s}}^{(1)}_{-1}&=&\frac{1}{\sqrt{2}}(\boldsymbol{\hat{s}}_x-i\boldsymbol{\hat{s}}_y),
    \\\nonumber
    \boldsymbol{\hat{s}}^{(1)}_{+1}|-\frac{1}{2}\rangle &=& -\frac{1}{\sqrt{2}}|+\frac{1}{2}\rangle,
    \\\nonumber
    \boldsymbol{\hat{s}}^{(1)}_{-1}|+\frac{1}{2}\rangle&=&\frac{1}{\sqrt{2}}|-\frac{1}{2}\rangle.
\end{eqnarray}

Now that we have all the preliminary tools, we will derive the matrix elements.
For the elements in the form of $\langle\mathcal{A}|H_{\text{ss}}|\mathcal{A}'\rangle$ and $\langle\mathcal{B}|H_{\text{ss}}|\mathcal{B}'\rangle$, according to Table \ref{tab:mat_el2}, only the parts of $H_{\text{ss}}$ that transform as IR $A_1$ would yield non-zero values. $|\mathcal{A}\rangle$ and $|\mathcal{A}'\rangle$ can be any of states in Table \ref{tab:configuration} that transform as IR $A_2$ and similarly $|\mathcal{B}\rangle$ and $|\mathcal{B}'\rangle$ can be any of states in Table \ref{tab:configuration} that transform as IR $B_2$. $\{\hat{\boldsymbol{D}}_{xx},\hat{\boldsymbol{D}}_{yy},\hat{\boldsymbol{D}}_{zz}\}$ are the only components of $\hat{\boldsymbol{D}}$ that transform as IR $A_1$. In the spherical tensor form, $\hat{\boldsymbol{D}}_{0}$ transforms as IR $A_1$ and first two components of $\hat{\boldsymbol{D}}_{\pm2}$ also transform as IR $A_1$. Therefore, just $\hat{\boldsymbol{D}}_{0}$, and $\hat{\boldsymbol{D}}_{\pm2}$ contribute to non-zero values. At last, according to Eq.~(\ref{eq:zero_tensor}), $\Delta S\in\{0,\pm2\}$.
Similarly, for $\langle\mathcal{A}|H_{\text{ss}}|\mathcal{B}\rangle$ according to Table \ref{tab:mat_el2}, only the parts of $H_{\text{ss}}$ that transform as IR $B_1$ would yield non-zero values. $\{\hat{\boldsymbol{D}}_{xz},\hat{\boldsymbol{D}}_{zx}\}$ are the only components of $\hat{\boldsymbol{D}}$ that transform as IR $B_1$. Therefore, just $\hat{\boldsymbol{D}}_{\pm1}$ contributes to non-zero values. For these matrix elements, according to Eq.~(\ref{eq:zero_tensor}), we have $\Delta S\in\{\pm1\}$.

\begin{equation}
\begin{split}
    \left.\begin{matrix}
        \langle\mathcal{A}|H_{\text{ss}}|\mathcal{A}'\rangle\\
        \langle\mathcal{B}|H_{\text{ss}}|\mathcal{B}'\rangle
    \end{matrix}\right\}
    &\xrightarrow{}
    A_1
    \xrightarrow{}
    \hat{\boldsymbol{D}}_{0},\ \hat{\boldsymbol{D}}_{\pm2}
    \xrightarrow{}
    \Delta S\in\{0,\pm2\}\\
        \langle\mathcal{A}|H_{\text{ss}}|\mathcal{B}\rangle
        &\xrightarrow{}
        B_1
        \xrightarrow{} \hat{\boldsymbol{D}}_{\pm1}
        \xrightarrow{}
        \Delta S\in\{\pm1\}
    \label{eq:spin_selection_rule}
\end{split}
\end{equation}

After doing all the calculations, we get the matrix elements for the spin-spin interaction, shown below.

\begin{widetext}
\begin{gather}
    H_\text{ss}= \frac{\mu_0\gamma_\text{e}^2\hbar^2}{16\pi}\times
    \label{eq:HSS}\\
    \begin{blockarray}{ccccccccccccccc}
        \qquad\qquad\qquad & \mathcal{A}^{0,d}_{+1/2} & \mathcal{A}^{0,d}_{-1/2} & \mathcal{B}^{1,d}_{+1/2} & \mathcal{B}^{1,d}_{-1/2} & \mathcal{B}^{2,d}_{+1/2} & \mathcal{B}^{2,d}_{-1/2} & \mathcal{A}^{3,q}_{+1/2} & \mathcal{A}^{3,q}_{-1/2} & \mathcal{A}^{3,q}_{+3/2} & \mathcal{A}^{3,q}_{-3/2} & \mathcal{A}^{3,d'}_{+1/2} & \mathcal{A}^{3,d'}_{-1/2} & \mathcal{A}^{3,d''}_{+1/2} & \mathcal{A}^{3,d''}_{-1/2}\\
        \begin{block}{c(cccccccccccccc)}
            \mathcal{A}^{0,d}_{+1/2} & 
            0\\
            \mathcal{A}^{0,d}_{-1/2} & 
            0 & 0\\
            \mathcal{B}^{1,d}_{+1/2} & 
            0 & 0 & 0\\
            \mathcal{B}^{1,d}_{-1/2} & 
            0 & 0 & 0 & 0\\
            \mathcal{B}^{2,d}_{+1/2} & 
            0 & 0 & 0 & 0 & 0\\
            \mathcal{B}^{2,d}_{-1/2} & 
            0 & 0 & 0 & 0 & 0 & 0\\
            \mathcal{A}^{3,q}_{+1/2} & 
            \mathcal{E}_1 & 0 & 0 & -\mathcal{F}_1 & 0 & \mathcal{F}_2 & -\mathcal{D}_0\\
            \mathcal{A}^{3,q}_{-1/2} & 
            0 & -\mathcal{E}_1 & \mathcal{F}_1 & 0 & -\mathcal{F}_2 & 0 & 0 & -\mathcal{D}_0\\
            \mathcal{A}^{3,q}_{+3/2} & 
            0 & -\mathcal{E}_2  & \frac{-1}{\sqrt{3}}\mathcal{F}_1 & 0 & \frac{1}{\sqrt{3}}\mathcal{F}_2 & 0 & 0 & \mathcal{E}_3 & \mathcal{D}_0\\
            \mathcal{A}^{3,q}_{-3/2} & 
            \mathcal{E}_2 & 0 & 0 & \frac{1}{\sqrt{3}}\mathcal{F}_1 & 0 & \frac{-1}{\sqrt{3}}\mathcal{F}_2 & \mathcal{E}_3 & 0 & 0 & \mathcal{D}_0\\
            \mathcal{A}^{3,d'}_{+1/2} & 0 & 0 & 0 & -\mathcal{G} & 0 & 0 & -\mathcal{H} & 0 & 0 & \mathcal{K}^* & 0\\
            \mathcal{A}^{3,d'}_{-1/2} & 0 & 0 & -\mathcal{G} & 0 & 0 & 0 & 0& -\mathcal{H} & \mathcal{K}^* & 0 & 0 & 0 \\
            \mathcal{A}^{3,d''}_{+1/2} & 0 & 0 & 0 & 0 & 0 & 0 & \mathcal{I} & 0 & 0 & \mathcal{L}^* & \mathcal{J} & 0 & 0\\
            \mathcal{A}^{3,d''}_{-1/2} & 0 & 0 & 0 & 0 & 0 & 0 & 0 & \mathcal{I} & \mathcal{L}^* & 0 & 0 & \mathcal{J} & 0 & 0\\
        \end{block}
    \end{blockarray}
    \notag
\end{gather}
\end{widetext}



The matrix elements used above are defined as

\begin{eqnarray}
    \mathcal{D}_0 &=& \frac{1}{2\sqrt{5}}\Big (\langle bb' - b'b
    |\hat{D}^{(2)}_{zz}|
    bb' - b'b \rangle
    \\&&\qquad\qquad\quad\nonumber
    +
    \langle ba-ab
    |\hat{D}^{(2)}_{zz}|
    ba-ab \rangle
    \\&&\qquad\qquad\quad\nonumber
    +
    \langle ab'-b'a
    |\hat{D}^{(2)}_{zz}|
    ab'-b'a \rangle \Big),
    \\\nonumber
    \mathcal{E}_1 &=& \frac{1}{15}\langle ab'-b'a
    |\hat{D}^{(2)}_{zz}|
    ab-ba \rangle,
    \\\nonumber
    \mathcal{E}_2 &=& \frac{1}{10}\langle ab'-b'a
    |\hat{D}^{(2)}_{xx}-\hat{D}^{(2)}_{yy}|
    ab-ba \rangle,
    \\\nonumber
    \mathcal{E}_3 &=&
    \frac{1}{\sqrt{30}}\Big (
    \langle ab'-b'a
    |\hat{D}^{(2)}_{xx}-\hat{D}^{(2)}_{yy}|
    ab'-b'a \rangle
    \\&&\qquad\qquad\quad\nonumber
    +
    \langle bb' - b'b
    |\hat{D}^{(2)}_{xx}-\hat{D}^{(2)}_{yy}|
    bb' - b'b \rangle
    \\&&\qquad\qquad\quad\nonumber
    +
    \langle ab-ba
    |\hat{D}^{(2)}_{xx}-\hat{D}^{(2)}_{yy}|
    ab-ba \rangle \Big ),
    \\\nonumber
    \mathcal{F}_1 &=& \frac{\sqrt{3}}{2\sqrt{5}}\langle bb'-b'b
    |\hat{D}^{(2)}_{xz}|
    ab-ba \rangle,
    \\\nonumber
    \mathcal{F}_2 &=& \frac{\sqrt{3}}{2\sqrt{5}}\langle ab'-b'a
    |\hat{D}^{(2)}_{xz}|
    bb'-b'b \rangle,\nonumber
\end{eqnarray}

\begin{eqnarray}
    \mathcal{G}&=&\frac{1}{\sqrt{30}}\langle bb'-b'b|\hat{D}^{(2)}_{xz}|ba-ab\rangle,\\
    \mathcal{H}&=&\frac{1}{6\sqrt{10}} \{
    \langle bb' - b'b
    |\hat{D}^{(2)}_{zz}|
    bb' - b'b \rangle\nonumber\\
    &&\qquad\qquad\quad
    +
    \langle ba - ab
    |\hat{D}^{(2)}_{zz}|
    ba - ab \rangle\nonumber\\
    &&\qquad\qquad\quad
    -2\langle ab' - b'a
    |\hat{D}^{(2)}_{zz}|
    ab' - b'a \rangle \},\nonumber\\
    \mathcal{I}&=&\frac{\sqrt{3}}{2\sqrt{10}}
    \{
    -\langle bb' - b'b
    |\hat{D}^{(2)}_{zz}|
    bb' - b'b \rangle\nonumber\\
    &&\qquad\qquad\quad
    +\langle ab' - b'a
    |\hat{D}^{(2)}_{zz}|
    ab' - b'a \rangle
    \},\nonumber\\
    \mathcal{J}&=&-\frac{1}{2\sqrt{15}}
    \langle ab' - b'a|\hat{D}^{(2)}_{zz}|ab' - b'a \rangle,\nonumber\\
    \mathcal{K}&=&\frac{1}{2\sqrt{15}}
    \{
    +\langle ab'-b'a
    |(\hat{D}^{(2)}_{xx}-\hat{D}^{(2)}_{yy})|
    ab'-b'a\rangle\nonumber\\
    &&\qquad\qquad\quad
    +\langle bb'-b'b
    |(\hat{D}^{(2)}_{xx}-\hat{D}^{(2)}_{yy})|
    bb'-b'b\rangle\nonumber\\
    &&\qquad\qquad\quad
    -2\langle ba-ab
    |(\hat{D}^{(2)}_{xx}-\hat{D}^{(2)}_{yy})|
    ba-ab\rangle\},\nonumber\\
    \mathcal{L}&=&\frac{1}{2\sqrt{5}}
    \{-\langle ab'-b'a
    |(\hat{D}^{(2)}_{xx}-\hat{D}^{(2)}_{yy})|
    ab'-b'a\rangle\nonumber\\
    &&\qquad\qquad\quad
    +\langle bb'-b'b
    |(\hat{D}^{(2)}_{xx}-\hat{D}^{(2)}_{yy})|
    bb'-b'b\rangle\}.\nonumber
\end{eqnarray}

\subsection{Dipole transitions}\label{sec:supp_dipole}
The dipole allowed transition rates would be proportional to the values defined below.

\begin{eqnarray}
    \mu_x &=& eE_x\langle a|x|b\rangle\\\nonumber
    \mu_x' &=& eE_x\langle a|x|b'\rangle\\\nonumber
    \mu_{z,0} &=& eE_z\{ \langle b|z|b\rangle + \langle a|z|a\rangle + \langle b'|z|b'\rangle\}\\\nonumber
    \mu_{z,1} &=& eE_z\{ 2\langle b|z|b\rangle + \langle a|z|a\rangle\}\\\nonumber
    \mu_{z,2} &=& eE_z\{ \langle b|z|b\rangle + 2\langle a|z|a\rangle\}\\\nonumber
    \mu_{z,3} &=& eE_z\{ 2\langle b|z|b\rangle + \langle b'|z|b'\rangle\}\\\nonumber
    \mu_{z}' &=& eE_z\langle b|z|b'\rangle
\end{eqnarray}

Furthermore, the matrix elements are as shown below.

\begin{widetext}
\begin{gather}
    H_\text{dipole}=
    \label{eq:dip} \\
    \begin{blockarray}{ccccccccccccccc}
        \qquad\qquad\qquad & \mathcal{A}^{0,d}_{+1/2} & \mathcal{A}^{0,d}_{-1/2} & \mathcal{B}^{1,d}_{+1/2} & \mathcal{B}^{1,d}_{-1/2} & \mathcal{B}^{2,d}_{+1/2} & \mathcal{B}^{2,d}_{-1/2} & \mathcal{A}^{3,q}_{+1/2} & \mathcal{A}^{3,q}_{-1/2} & \mathcal{A}^{3,q}_{+3/2} & \mathcal{A}^{3,q}_{-3/2} & \mathcal{A}^{3,d'}_{+1/2} & \mathcal{A}^{3,d'}_{-1/2}& \mathcal{A}^{3,d''}_{+1/2} & \mathcal{A}^{3,d''}_{-1/2}\\
        \begin{block}{c(cccccccccccccc)}
            \mathcal{A}^{0,d}_{+1/2} & 
            \mu_{z,1}
            \\
            \mathcal{A}^{0,d}_{-1/2} & 
            0 & \mu_{z,1}
            \\
            \mathcal{B}^{1,d}_{+1/2} & 
            -\mu_x & 0 & \mu_{z,2}
            \\
            \mathcal{B}^{1,d}_{-1/2} & 
            0 & \mu_x & 0 & \mu_{z,2}
            \\
            \mathcal{B}^{2,d}_{+1/2} & 
            \mu_x'^* & 0 & 0 & 0 & \mu_{z,3}
            \\
            \mathcal{B}^{2,d}_{-1/2} & 
            0 & \mu_x'^* & 0 & 0 & 0 & \mu_{z,3}
            \\
            \mathcal{A}^{3,q}_{+1/2} & 
            0 & 0 & 0 & 0 & 0 & 0 & \mu_{z,0}
            \\
            \mathcal{A}^{3,q}_{-1/2} & 
            0 & 0 & 0 & 0 & 0 & 0 & 0 & \mu_{z,0}
            \\
            \mathcal{A}^{3,q}_{+3/2} & 
            0 & 0 & 0 & 0 & 0 & 0 & 0 & 0 & \mu_{z,0}
            \\
            \mathcal{A}^{3,q}_{-3/2} & 
            0 & 0 & 0 & 0 & 0 & 0 & 0 & 0 & 0 & \mu_{z,0}\\
            \mathcal{A}^{3,d'}_{+1/2} & \frac{3\mu'^*_z}{\sqrt{6}} & 0 & -\frac{3\mu'^*_x}{\sqrt{6}} & 0 & 0 & 0 & 0 & 0 & 0 & 0 & 0\\
            \mathcal{A}^{3,d'}_{-1/2} & 0 & \frac{3\mu'^*_z}{\sqrt{6}} & 0 & -\frac{3\mu'^*_x}{\sqrt{6}} & 0 & 0 & 0 & 0 & 0 & 0 & 0 & 0 \\
            \mathcal{A}^{3,d''}_{+1/2}  & -\frac{\mu'^*_z}{\sqrt{2}} & 0 & -\frac{\mu'^*_x}{\sqrt{2}} & 0 & \frac{2\mu'_x}{\sqrt{2}} & 0 & 0 & 0 & 0 & 0 & 0 & 0 & 0\\
            \mathcal{A}^{3,d''}_{-1/2} & 0 & -\frac{\mu'^*_z}{\sqrt{2}} & 0 & -\frac{\mu'^*_x}{\sqrt{2}} & 0 & \frac{2\mu'_x}{\sqrt{2}} & 0 & 0 & 0 & 0 & 0 & 0 & 0 & 0\\
        \end{block}
    \end{blockarray}
    \notag
\end{gather}
\end{widetext}


\section{Magnetic interaction}\label{sec:sup_HB}
We define the following values for single molecular orbitals.
\begin{equation}
\begin{split}
    \eta &= \langle a|l_y|b\rangle\\
    \eta' &= \langle a|l_y|b'\rangle 
\end{split}
\end{equation}

{\scriptsize
\begin{widetext}
\begin{gather}
    H_\text{B}= \frac{\gamma_\text{e}\hbar}{2}\times 
    \label{eq:magnetic} \\
    \begin{blockarray}{ccccccccccc}
        \qquad\qquad\qquad & \mathcal{A}^{0,d}_{+1/2} & \mathcal{A}^{0,d}_{-1/2} & \mathcal{B}^{1,d}_{+1/2} & \mathcal{B}^{1,d}_{-1/2} & \mathcal{B}^{2,d}_{+1/2} & \mathcal{B}^{2,d}_{-1/2} & \mathcal{A}^{3,q}_{+1/2} & \mathcal{A}^{3,q}_{-1/2} & \mathcal{A}^{3,q}_{+3/2} & \mathcal{A}^{3,q}_{-3/2}\\
        \begin{block}{c(cccccccccc)}
            \mathcal{A}^{0,d}_{+1/2} & 
            B_z
            \\
            \mathcal{A}^{0,d}_{-1/2} & 
            (B_x+iB_y) & -B_z
            \\
            \mathcal{B}^{1,d}_{+1/2} & 
            \frac{2B_y}{g_\text{e}}\eta & 0 & B_z
            \\
            \mathcal{B}^{1,d}_{-1/2} & 
            0 & \frac{2B_y}{g_\text{e}}\eta & (B_x+iB_y) & -B_z
            \\
            \mathcal{B}^{2,d}_{+1/2} & 
            \frac{2B_y}{g_\text{e}}\eta'^* & 0 & 0 & 0 & B_z
            \\
            \mathcal{B}^{2,d}_{-1/2} & 
            0 & \frac{2B_y}{g_\text{e}}\eta'^* & 0 & 0 & (B_x+iB_y) & -B_z
            \\
            \mathcal{A}^{3,q}_{+1/2} & 
            0 & 0 & 0 & 0 & 0 & 0 & B_z
            \\
            \mathcal{A}^{3,q}_{-1/2} & 
            0 & 0 & 0 & 0 & 0 & 0 & 2(B_x+iB_y) & -B_z
            \\
            \mathcal{A}^{3,q}_{+3/2} & 
            0 & 0 & 0 & 0 & 0 & 0 & \sqrt{3}(B_x-iB_y) & 0 & 3B_z
            \\
            \mathcal{A}^{3,q}_{-3/2} & 
            0 & 0 & 0 & 0 & 0 & 0 & 0 & \sqrt{3}(B_x+iB_y) & 0 & -3B_z\\
        \end{block}
    \end{blockarray}
    \notag
\end{gather}
\end{widetext}}

\section{Hyperfine interaction}\label{sec:sup_hyperfine}

We can write the hyperfine interaction in the form of spherical components as below.

\begin{equation}
\begin{split}
    \hat{V}_\text{mhf}&=-C_\text{mhf}\sum_i[\hat{J}_{i}^{(2)}\otimes\hat{A}_{i}^{(2)}]^{(0)}\\
        &=-\frac{C_\text{mhf}}{\sqrt{5}}\sum_i\sum_{q=-2}^{q=+2}(-1)^{2-q}J_{i,-q}^{(2)}A_{i,+q}^{(2)},
    \label{eq:appendix_Vmhf}
\end{split}
\end{equation}

After doing the calculations, we end up with the following Hamiltonian for the ground state.

\begin{gather}
    V_\text{mhf}= -\frac{C_\text{mhf}}{12\sqrt{5}}
    \begin{blockarray}{cccccc}
        \qquad\qquad & \Psi_{1} & \Psi_{2} & \Psi_{3} & \Psi_{4} & \;\\
        \begin{block}{c(cccc)c}
            \Psi_{1} & G_0
            \\
            \Psi_{2} & 
            0 & -2G_0
            \\
            \Psi_{3} & 
            -3\sqrt{2}G_1 & 0 & G_0
            \\
            \Psi_{4} & 
            0 & -G_0 & 0 & 0
            \\
        \end{block}
    \end{blockarray},
    \label{eq:Vmhf}
\end{gather}

where
\begin{equation}
\begin{split}
    G_0 &= \langle a
    |(2A_{zz}^{(2)}-A_{xx}^{(2)}-A_{yy}^{(2)})|
    a\rangle,\\
    G_1 &= \langle a
    |(A_{xx}^{(2)}-A_{yy}^{(2)})|
    a\rangle.
\end{split}
\end{equation}

\begin{table}[htp]
    \centering
    \begin{tabular}{c c c}
        \multicolumn{3}{c}{Hyperfine interaction}\\
        \hline
        Eigenvalues & Eigenstates & Eigenstates in primary basis \\
        $-(1+\sqrt{2})G_0$ & $(1+\sqrt{2})\Psi_2+\Psi_4$ & $(1+\sqrt{2})|b\bar{b}a-\rangle+|b\bar{b}\bar{a}+\rangle$\\
        $-(1-\sqrt{2})G_0$ & $(1-\sqrt{2})\Psi_2+\Psi_4$ & $-|b\bar{b}a-\rangle+(1+\sqrt{2})|b\bar{b}\bar{a}+\rangle$\\
        $G_0-3\sqrt{2}G_1$ & $\Psi_1+\Psi_3$ & $|b\bar{b}a+\rangle+|b\bar{b}\bar{a}-\rangle$\\
        $G_0+3\sqrt{2}G_1$ & $-\Psi_1+\Psi_3$ & $-|b\bar{b}a+\rangle+|b\bar{b}\bar{a}-\rangle$\\
    \end{tabular}
    \caption{Eigensystem of the hyperfine interaction for the ground state in Eq.~(\ref{eq:Vmhf}). Eigenvalues should be multiplied by $-\frac{C_\text{mhf}}{12\sqrt{5}}$.}
    \label{tab:Vmhf_eigensystem} 
\end{table}

\section{Calculations of the multi-configuration states\label{sec:singlet-triplet}} 
Here, we use the single-configuration states to estimate the energies of the corresponding multi-configuration states. We only look at the spin-up quartet and doublet states, but the spin-down calculations are similar. We start with quartet and doublet superposition states.

\begin{equation}
\begin{split}
    |\Psi_q\rangle&=\frac{1}{\sqrt{3}}|\beta\alpha\alpha+\alpha\beta\alpha+\alpha\alpha\beta\rangle\\
    |\Psi_{d'}\rangle&=\frac{1}{\sqrt{6}}|\beta\alpha\alpha+\alpha\beta\alpha-2\alpha\alpha\beta\rangle\\
    |\Psi_{d''}\rangle&=\frac{1}{\sqrt{2}}|-\beta\alpha\alpha+\alpha\beta\alpha\rangle\\
\end{split}
\end{equation}

\begin{table}[h]
    \centering
    \begin{tabular}{c|c}
        state & energy\\
        \hline
        $|\alpha\alpha\beta\rangle$ & 4.7 eV \\ $|\beta\alpha\alpha\rangle$ & 5.4 eV \\ $|\alpha\beta\alpha\rangle$ & 5.5 eV
    \end{tabular}
    \caption{Electron spin configurations corresponding to $|bab'\rangle$ and their energies obtained from DFT. $\alpha \; (\beta)$ represents spin up (down).}
    \label{tab:single_energy}
\end{table}

Since $\ket{\alpha\alpha\alpha}$ is also a quartet state, we have $E_q\equiv E[|\Psi_q\rangle]=E[|\alpha\alpha\alpha\rangle]$. By using the equation
\begin{equation}
\begin{split}
    \frac{1}{\sqrt{3}}(|\Psi_q\rangle-\sqrt{2}|\Psi_{d'}\rangle)=|\alpha\alpha\beta\rangle,
\end{split}
\end{equation}
we have $E[|\alpha\alpha\beta\rangle]=\frac{1}{3}(E_q+2E[|\Psi_{d'}\rangle])$. Thus, $E_{d'}\equiv E[|\Psi_{d'}\rangle]=(3E[|\alpha\alpha\beta\rangle]-E_q)/2$. Based on our DFT calculations in Table \ref{tab:single_energy}, $E[|\alpha\alpha\beta\rangle]=4.7$ eV and $E_q=4.1$ eV, which implies that $E_{d'}=(3\times4.7-4.1)/2=5$ eV. Next, we define auxiliary state $\phi$:
\begin{equation}
\begin{split}
    |\phi\rangle&\equiv\frac{1}{2\sqrt{3}}(\sqrt{3}|\Psi_{d'}\rangle+3|\Psi_{d''}\rangle)\\
    &=\frac{1}{2\sqrt{6}}| - 2\beta\alpha\alpha+4\alpha\beta\alpha - 2\alpha\alpha\beta\rangle\\
    &=\frac{1}{\sqrt{6}}| - \beta\alpha\alpha + 2\alpha\beta\alpha - \alpha\alpha\beta\rangle
\end{split}
\end{equation}
By using this auxiliary state, we show that
\begin{equation}
\begin{split}
    \frac{1}{\sqrt{3}}(|\Psi_q\rangle+\sqrt{2}|\phi\rangle)=|\alpha\beta\alpha\rangle.
\end{split}
\end{equation}
Similar to previous calculation, we can show that
\begin{equation}
\begin{split}
    E[|\alpha\beta\alpha\rangle] &= \frac{1}{3}(E_q+2E_\phi)\\
    &=\frac{1}{3}\Bigg\{E_q+\frac{1}{6}(3E_{d'}+9E_{d''}+3\sqrt{3}\langle\Psi_{d'}|H|\Psi_{d''}\rangle\\
    &\qquad
    +3\sqrt{3}\langle\Psi_{d''}|H|\Psi_{d'}\rangle)\Bigg\}\\
    &=\frac{1}{6}\left\{2E_q+E_{d'}+3E_{d''}\right\}.
\end{split}
\end{equation}
Therefore, we can calculate $E_{d''}$ as below.
\begin{equation}
\begin{split}
    E_{d''}&=\frac{1}{3}\left\{6E[|\alpha\beta\alpha\rangle]-2E_q-E_{d'}\right\}\\
    &=\frac{1}{3}(6\times5.5-2\times4.1-5)=6.6 \; \text{eV}
\end{split}
\end{equation}

\section{ODMR signal\label{sec:sup_ODMR}}
\begin{figure}[H]
    \centering
    \includegraphics[width=6.5cm]{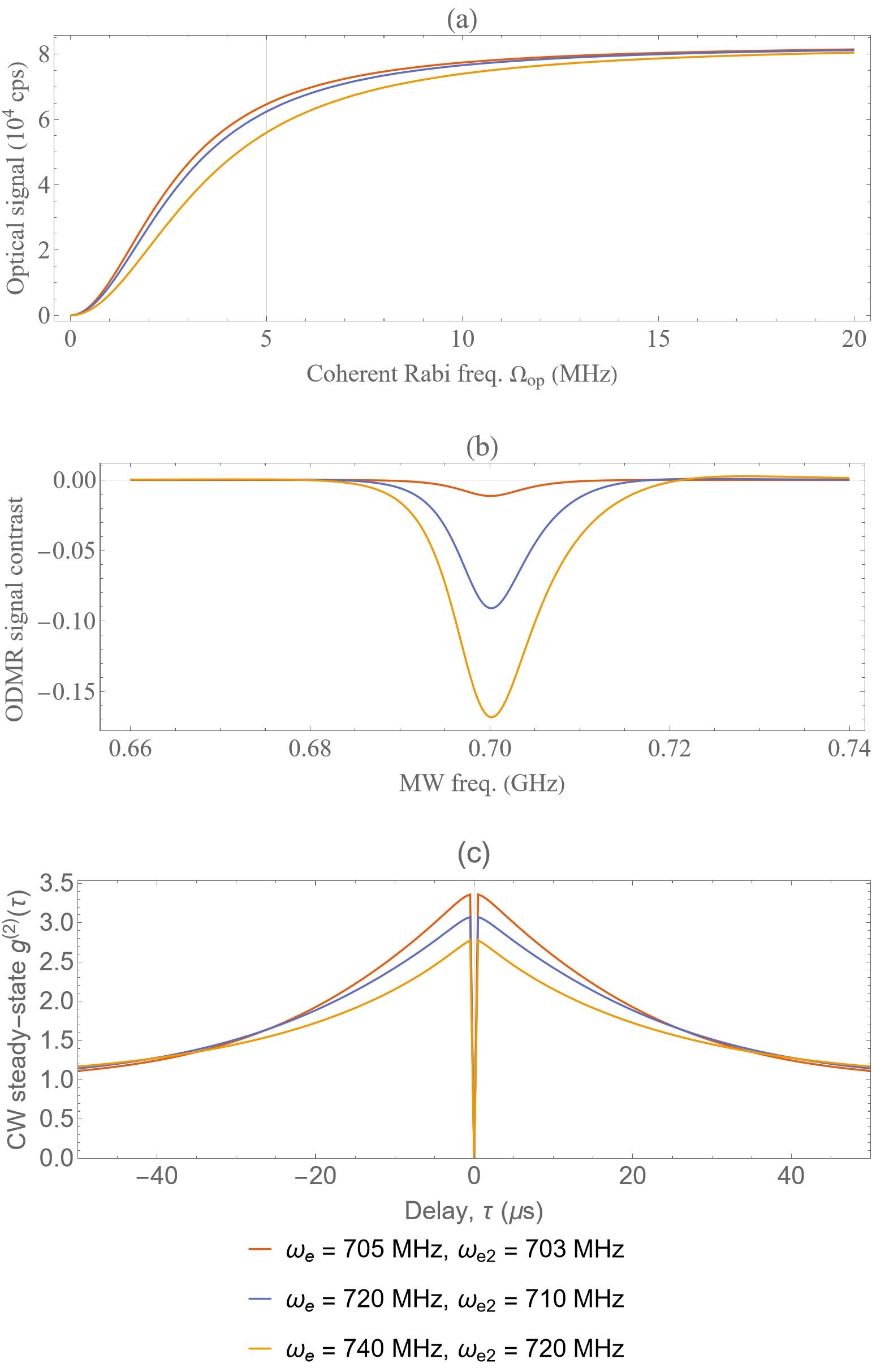}
    \caption{The effect of changing $\omega_\text{e}$ and $\omega_\text{e2}$ on the optical signal, ODMR contrast, and the second-order correlation function. The grid line shows the Rabi frequency at 5 MHz.}
    \label{fig:resonance}
\end{figure}

\begin{figure}[H]
    \centering
    \includegraphics[width=6.5cm]{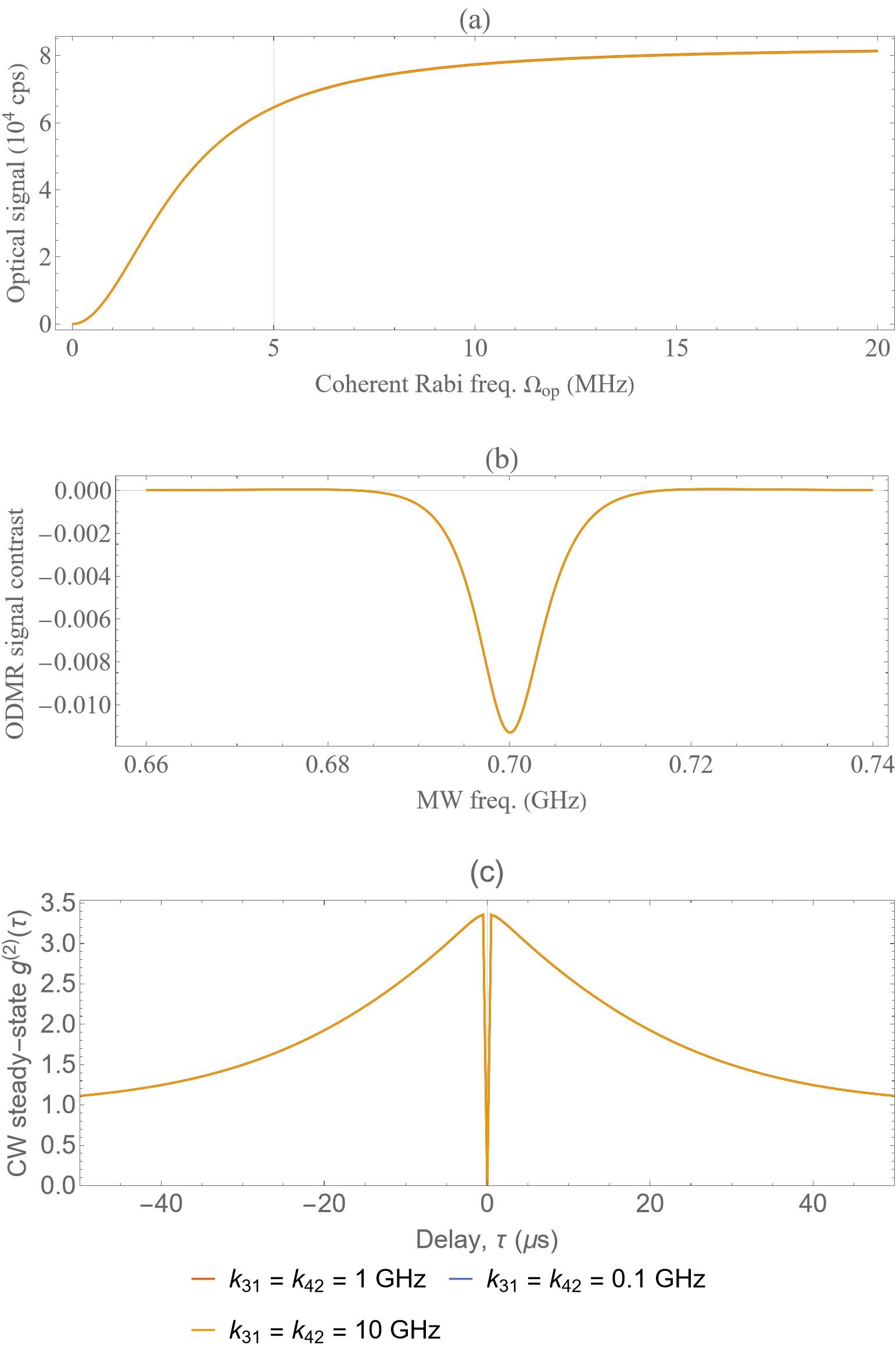}
    \caption{The effect of changing $k_{31}$ and $k_{42}$ on the optical signal, ODMR contrast, and the second-order correlation function.}
    \label{fig:k31=k42}
\end{figure}

In this section, we study the effect of changing parameters used for ODMR signal in Table \ref{tab:ODMR_par}. In each of the following figures, we change only one or two parameters to see their effect on our model. In Fig. \ref{fig:resonance}, we checked the effect of changing the first and the second excited state spin splittings $\omega_e$ and $\omega_{e2}$. The result shows that although changing them would affect the magnitude of the ODMR signal, there is only one resonance in the ODMR signal, and it is because of the ground-state spin splitting $\omega_g$. This might be due to the high decay rates from the excited state compared to microwave driving, which makes the spin-flip rates in the excited state negligible compared to the ground state. All lines in Fig. \ref{fig:k31=k42} overlap indicating that changing $k_{31}$ and $k_{42}$ does not change the results. This is expected because the meta-stable and dark states, which have important roles in seeing the ODMR signal, are higher in energy than the first excited state. Therefore, changing the decay rates from the first excited state to the ground state does not affect the results. 

\begin{figure}[H]
    \centering
    \includegraphics[width=6.5cm]{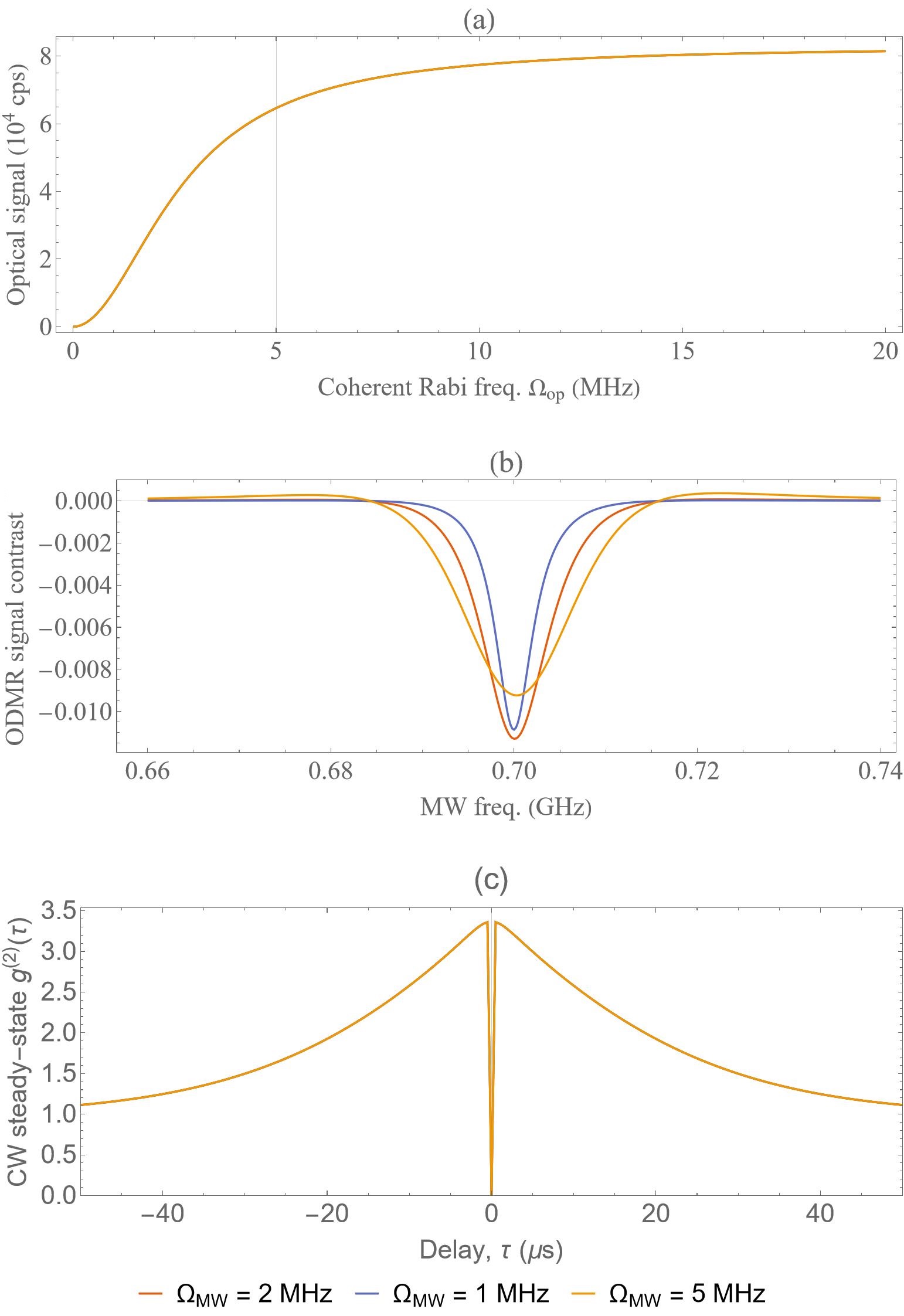}
    \caption{The effect of changing $\Omega_\text{MW}$ on the optical signal, ODMR contrast, and the second-order correlation function. The grid line shows the Rabi frequency at 5 MHz.}  
    \label{fig:Omega_MW}
\end{figure}

\begin{figure}[H]
    \centering
    \includegraphics[width=6.5cm]{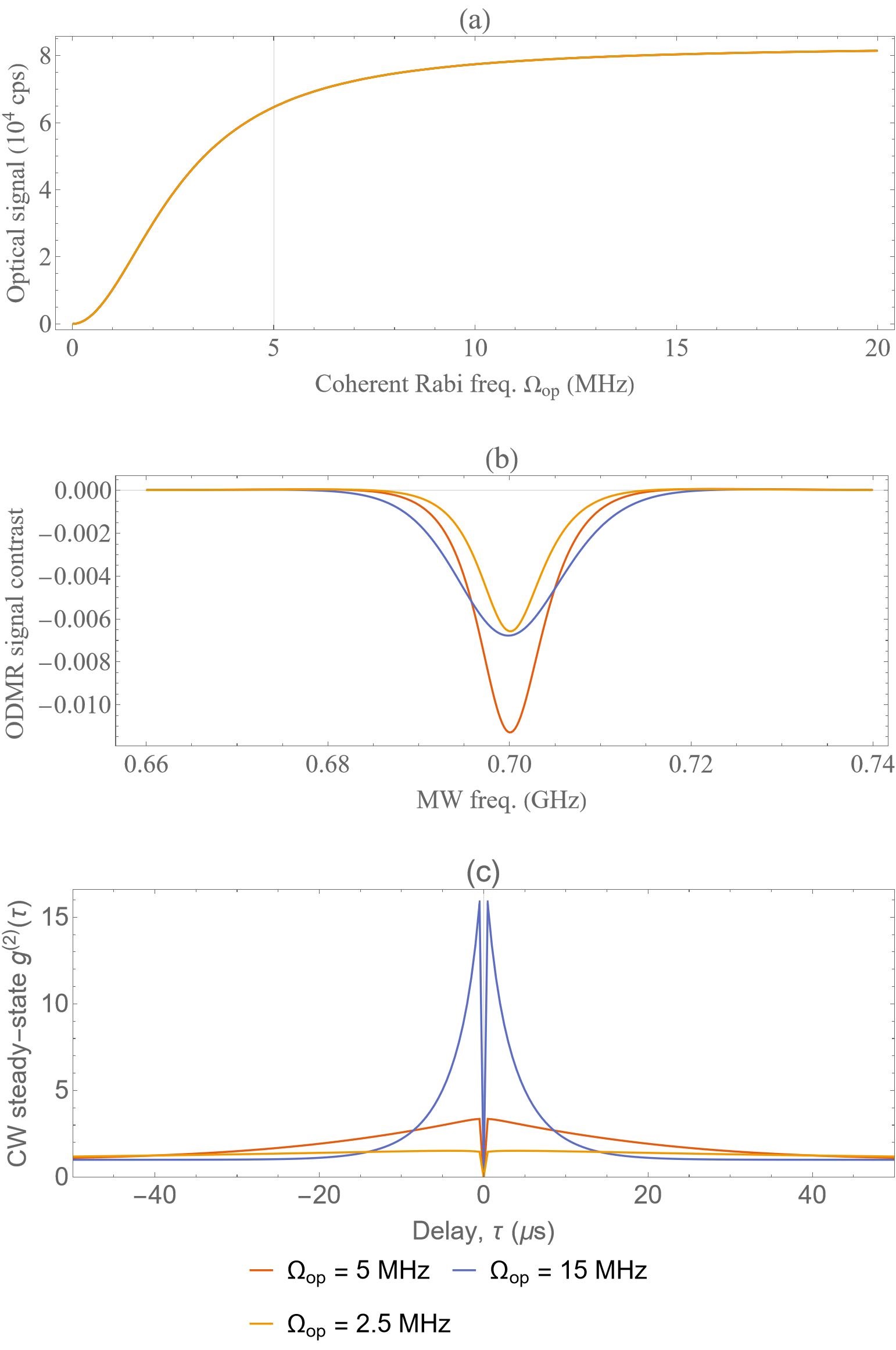}
    \caption{The effect of changing $\Omega_\text{op}$ on the optical signal, ODMR contrast, and the second-order correlation function. The grid line shows the Rabi frequency at 5 MHz.}
    \label{fig:Omega_op}
\end{figure}

\begin{figure}[H]
    \centering
    \includegraphics[width=6.5cm]{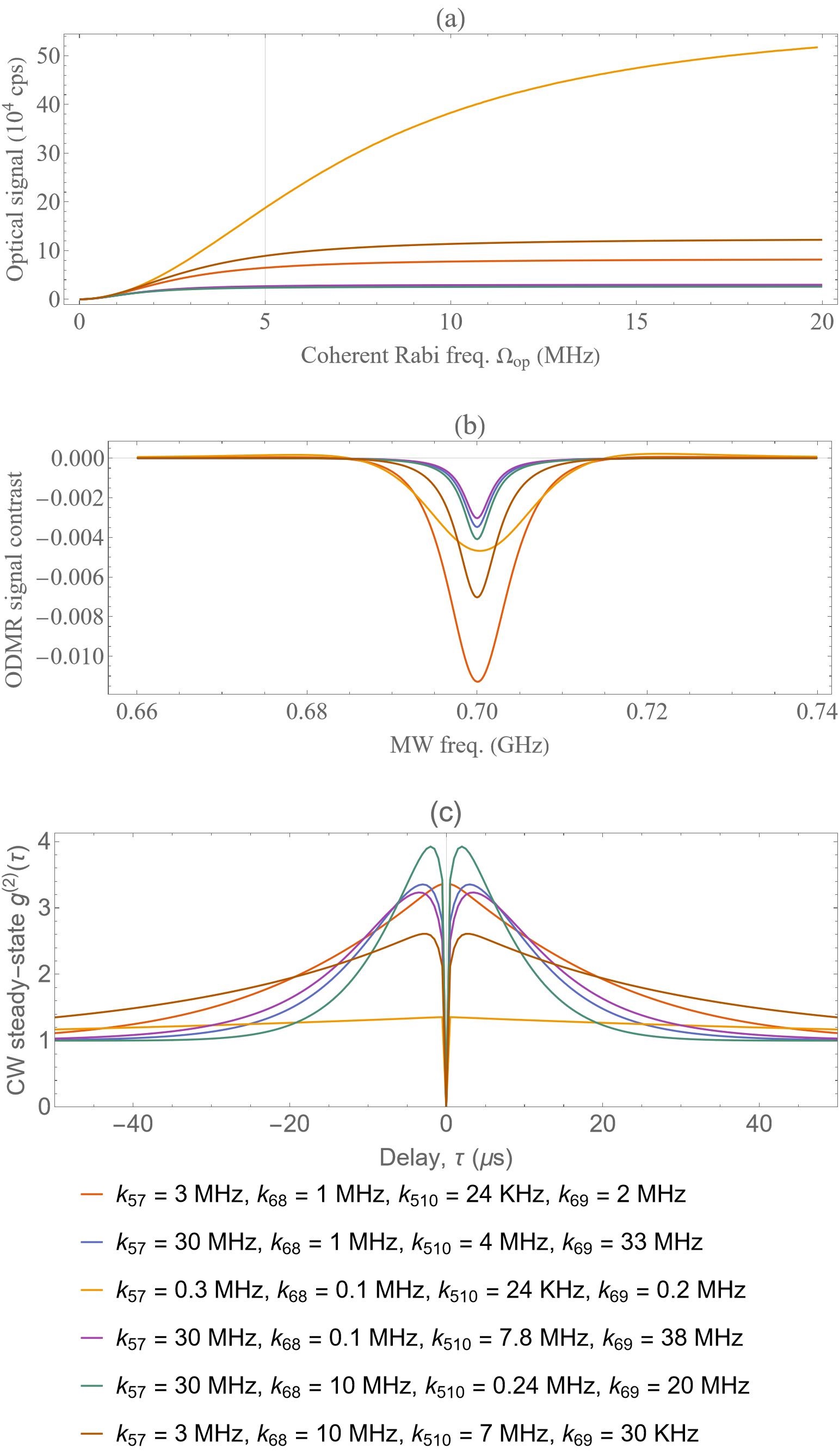}
    \caption{The effect of changing $k_{57}$ and $k_{68}$ on the optical signal, ODMR contrast, and the second-order correlation function. The grid line shows the Rabi frequency at 5 MHz. Based on the matrix elements of the spin-orbit and spin-spin interactions, the $k_{57}$ and $k_{68}$ rates are related to the $k_{510}$ and $k_{69}$ rates. So changing each of them will affect the other two.}
    \label{fig:k57 & k68}
\end{figure}


\bibliographystyle{unsrt}
\bibliography{references}

\end{document}